# Advances in the theory and methods of computational vibronic spectroscopy

Sergey A. Astakhov*, Victor I. Baranov† and Lev A. Gribov†

\* *Department of Chemistry and Biochemistry, Utah State University, Logan, UT 84322, USA*
( *s-astakhov@yandex.ru* , *http://www.astakhov.newmail.ru* )

† *Vernadsky Institute of Geochemistry and Analytical Chemistry, Russian Academy of Sciences, Kosygin str. 19, 119991 Moscow, Russia*
( *baranov@geokhi.ru* , *gribov@geokhi.ru* )

**Abstract**

We discuss semiempirical approaches and parametric methods developed for modeling molecular vibronic spectra. These methods, together with databases of molecular fragments, have proved efficient and flexible for solving various problems ranging from detailed interpretation of conventional vibronic spectra and calculation of radiative transition probabilities to direct simulations of dynamical (time-resolved) spectra and spectrochemical analysis of individual substances and mixtures. A number of specific examples and applications presented here show the potential of the semiempirical approach for predictive calculations of spectra and solution of inverse spectral problems. It is noteworthy that these advances provide computational insights into developing theories of photoinduced isomer transformations and nonradiative transitions in polyatomic molecules and molecular ensembles, theory of new methods for standardless quantitative spectral analysis.

**Introduction**

Polyatomic molecules, supramolecules and molecular associates can exchange information and transfer energy through collisions and chemical reactions, and also through elementary acts such as absorption, emission or scattering of electromagnetic waves. Beyond trivial absorption and consecutive spontaneous emission, these rather complicated processes may result in and be used for irreversible photochemical transformations of molecular structure; creating an inverted population, data storage; and pattern recognition [1-4] at the molecular level. These peculiarities of objects from the world of molecules make this world very attractive for studies and design of new materials or drugs, molecular devices [5-9], fundamentally new electronic components for computers and communication systems.

However, it is quite clear that purely empirical work along these lines, even if supported by brute-force *ab initio* calculations, will be difficult or even impossible. This is due to the sheer size and high dimensionality of realistic systems. It therefore seems reasonable that further progress will rely heavily on investigative computer simulations and experiments. A number of easily adjustable molecular models and computational methods come from the theory of vibrational and vibronic spectra whose "speciality" is basically finding energy levels and transition probabilities for complex molecular systems subjected to various perturbations and external fields. For this reason, advances in theoretical molecular spectroscopy and its methods, which may seem far from being able to describe processes any more complicated than basic absorption or emission of radiation, ultimately determine and spur the development of other closely related chapters, such as quantum chemistry of large (bio)molecules. Further, and this is more important, prognostic capabilities of the theory of spectra can be most easily tested experimentally – in this sense simulations of spectra give one the shortest bridge from molecular models to observed quantities; for this reason theoretical spectroscopy has been long used in studies of structure and reactivity of polyatomic molecules in their ground and excited states.

As is believed natural for any theoretical generalization claiming to provide a description of complex systems, practical applications immediately manifest the principle of complementarity introduced by Niels Bohr. Theory of spectra originates from rigorous quantum mechanical concepts and approaches, but from the other hand — it relies on an appropriate choice and parameterization of a particular molecular model that inevitably implies semi-empirical treatment of a problem at hand. This occurs primarily in pursuing clarity when one wishes to develop a language for describing complex objects. But still the other side of the coin is that a purely "mathematical solution" might give results that would be hard to interpret and compare with an experiment. A good illustration comes from quantum chemistry where one just cannot find rigorous proofs for some fundamental concepts ranging from convergence of the self-consistent field procedure, the idea of LCAO all the way down to the choice of Gaussian basis sets.

A comprehensive approach to the problems in the theory of spectra with special emphasis paid to practical aspects has been pursued in a series of monographs by one of the authors and his colleagues [10-17]. This article is intended as a sequel and further development of the circle of ideas introduced there.

Here we will use the Born-Oppenheimer adiabatic approximation; hence the vibrational structure of a spectrum can be understood, e.g., quantitatively in terms of differences between ground state and excited state potential energy surfaces (PES). Significant progress in this field gave rise to new quite accurate and efficient methods for computing all the matrix elements [16, 18-20].

Thus, the central problem of adiabatic theory of spectra reduces now to the problem of how to obtain changes in geometry and force constants occurring upon electronic excitation. It is worth noting here that for polyatomic molecules these differences in PES are typically small (of the order of several percent), so appropriate methods should be employed to achieve spectroscopic accuracy.

Existing approaches make it possible to perform direct calculations of spectra for adiabatic molecular models, achieving good quantitative agreement with experiment (see, for example, [14, 15, 21-24]). This is not only a window into interpretation of spectral experiment, but it also leads to refinement of models through solution of inverse problems [25].

However, the level of adiabatic theory that has been achieved so far, and questions that have arisen, suggest that further development is necessary, perhaps, mainly with the aim to improve the system of parameters and extend computational methods in accord with experiments. Here we refer primarily to high resolution laser spectroscopy in supersonic jets, experimental techniques based on dispersed single vibronic level fluorescence spectra [26-38] and, especially, the exponentially growing field of time-resolved spectroscopy with closely related modern femto-sciences — femtochemistry and femtobiology [39-46]. All this provides strong motivation for the development of corresponding theoretical methods and predictive molecular models.

## 1. The system of parameters for adiabatic molecular model in the theory of vibronic spectra

As the experimental techniques have developed and new methods came into play, the role of detailed, accurate and, more importantly, predictive calculations of vibronic spectra have become principle in the studies of structural and spectroscopic properties of polyatomic molecules. Due to quite general reasons, but not just because of lack of computational resources, much progress in this field is to be expected from parametric semiempirical methods supplied with databases of molecular fragments and physically meaningful systems of parameters. The possibility of such theory for adiabatic molecular models was first discussed in [47] and found further development in [48-50]. The basic idea is to describe PES of the molecule in terms of two groups of parameters — first and second derivatives of Coulomb and resonance one-electron integrals in the basis of hybridized atomic orbitals (HAO) with respect to natural coordinates (the first derivatives are responsible for changes in geometry of a molecule, while the second — for changes in force constants). The important point also is that these parameters are the same for different electronic states.

In the framework of the parametric approach, most of analytical expressions for changes in PES upon electronic excitation can be significantly simplified by retaining only the leading diagonal terms. This along with characteristic structure of changes in electron density matrix, pronounced local properties of the derivatives and ranking by their absolute values, makes the number of these parameters just small, whereas transferability of parameters in homologous series and optimal semiempirical choice of their specific values for certain molecular fragments have provided a reasonable basis for direct calculations of one-photon absorption and fluorescence vibronic spectra and properties of excited states. The use of molecular fragments is essential, providing that calibration of model parameters can be done for small molecules and one type of experiment (conventional absorption or fluorescence). Models of complex compounds can then be built from these fragments (pre-stored in databases [48-53]) and used for simulations of spectra either of the same or of the other type. Such approach has been applied to extend existing parametric methods to the case of time-resolved vibronic spectra, which — having one extra dimension— contain more spectroscopic information. We shall address this issue and summarize our results on direct simulations of dynamical spectra in section 3.

In adiabatic approximation, models of combining electronic states can be described by the parameters of PES, namely by the location of its minimum (equilibrium geometry, $s$) and curvature (force constants, $u$). This system of parameters have proved suitable in the theory of vibrations and vibrational spectra of polyatomic molecules, since it gives clear physical picture and holds some transferability in series of molecules and structural groups. In turn, vibrational structure of electronic spectra is related to changes in PES upon excitation ($\Delta s$, $\Delta u$) which for large molecules are typically small (of the order of 5%) unless transitions between isomers are considered. Correspondingly, required transferability of these quantities must be at least 2 orders of magnitude higher than that of parameters of the ground state ($s^{(0)}$, $u^{(0)}$). However, calculations show [54] that even for polyenes, that have rather similar geometrical and electronic structures, straightforward transfer of $\Delta s$ and $\Delta u$ from one molecule to another results in serious errors. Besides, these values differ from one electronic states to another. It obviously raises severe difficulties on the way toward developing databases of molecular fragments for spectral calculations. Therefore, a theory of vibronic spectra should be based on an alternative system of "hidden" parameters of adiabatic molecular model that necessarily meet the following criteria:

(i) clear physical meaning and correlation with PES parameters;
(ii) locality that could allow to separate out characteristic structural groups (fragments);
(iii) transferability between similar species containing typical fragments;
(iv) suitability for ranking and choosing relatively small set of most relevant parameters;
(v) independence from small changes of electron density distribution in fragment upon excitation that allow to construct unified system of parameters for different excited states;

In developing the parametric semiempirical approach we will proceed from the methods proposed for calculation of parameters of excited state PES [55] and parameterization of molecular models [47, 56, 57].

In harmonic approximation, PES of the *n*-th electronic state expressed in terms of independent normal coordinates $Q^{(0)}$ of the ground state represents as:

$$E^{(n)} = E_0^{(n)} + \tilde{a}^{(n)} Q^{(0)} + \frac{1}{2} \tilde{Q}^{(0)} A^{(n)} Q^{(0)}, \qquad (1.1)$$

where $E_0^{(n)}$ is the energy of the *n*-th state at the equilibrium geometry ($Q^{(0)}=0$).

For the ground state we have

$$E^{(0)} = E_0^{(0)} + \frac{1}{2} \tilde{Q}^{(0)} A^{(0)} Q^{(0)} = E_0^{(0)} + \frac{1}{2} \tilde{Q}^{(0)} \Lambda Q^{(0)}, \qquad (1.2)$$

where the matrix of squares of vibrational frequencies $\Lambda$, as well as $Q^{(0)}$, for the ground state can be found by the standard methods in the theory of vibrations and IR spectra of polyatomic molecules [13]   According to the Hellmann-Feynman theorem, generalized forces $a^{(n)}$ (taken with opposite sign) associated with normal coordinates $Q^{(0)}$ have the form:

$$a_k^{(n)} = \frac{\partial E^{(n)}}{\partial Q_k^{(0)}} = \int \varphi_n^* \frac{\partial \hat{H}}{\partial Q_k^{(0)}} \varphi_n dv, \qquad (1.3)$$

or, in matrix notation:

$$a_k^{(n)} = Sp \left[ P^{(n)} \frac{\partial H_e}{\partial Q_k^{(0)}} \right], \qquad (1.4)$$

where $P^{(n)}$ is the density matrix in the *n*-th electronic state. In CNDO approximation $H_e = H + S\,V_{NN}$, with $H$ being the matrix of Coulomb and resonance one-electron integrals, $S$ – overlap matrix in AO basis, $V_{NN} = \sum_{\alpha,\beta} (Z_\alpha Z_\beta / r_{\alpha,\beta})$ – potential energy of repulsion between nuclei. The matrix of second derivatives reads:

$$A_{kl}^{(n)} = Sp\left[ P^{(n)} \frac{\partial^2 H_e}{\partial Q_k^{(0)} \partial Q_l^{(0)}} \right] + \frac{1}{2} Sp\left[ \frac{\partial P^{(n)}}{\partial Q_k^{(0)}} \frac{\partial H_e}{\partial Q_l^{(0)}} + \frac{\partial P^{(n)}}{\partial Q_l^{(0)}} \frac{\partial H_e}{\partial Q_k^{(0)}} \right]. \qquad (1.5)$$

Using (1.1) and relation between normal and natural coordinates of the ground state $q^{(0)} = L_q^{(0)} Q^{(0)}$, one can find changes in geometry (bond lengths, valence angles, etc.) upon excitation [55]:

$$\Delta s^{(n)} = s^{(n)} - s^{(0)} = -L_q^{(0)} \left( A^{(n)} \right)^{-1} a^{(n)}. \qquad (1.6)$$

Here it is assumed that the same but shifted system of natural coordinates can be used for excited states, which is reasonable for small transformations of molecular structure.

Since $a^{(0)} = 0$ for the ground state, (1.6) takes the form:

$$\Delta s^{(n)} = -L_q^{(0)} \left(A^{(n)}\right)^{-1} \Delta a^{(n)}, \tag{1.7}$$

where

$$\Delta a^{(n)} = a^{(n)} - a^{(0)} = Sp\left[\Delta P^{(n)} \frac{\partial H_e}{\partial Q_k^{(0)}}\right] = \widetilde{L}_q^{(0)} Sp\left[\Delta P^{(n)} \frac{\partial H_e}{\partial q_k^{(0)}}\right] \tag{1.8}$$

and $\Delta P^{(n)} = P^{(n)} - P^{(0)}$ is the matrix of changes in electron density upon excitation.

If we transform now (1.1) to natural coordinates $q^{(n)} = q^{(0)} - \Delta s^{(n)}$ and note that linear term should vanish, then for the force constants of excited state model $u_{kl}^{(n)}$ and their changes $\Delta u_{kl}^{(n)}$ we obtain:

$$E^{(n)} = E_{\min}^{(n)} + \frac{1}{2}\widetilde{q}^{(n)} L_p^{(0)} A^{(n)} \widetilde{L}_p^{(0)} q^{(n)}, \tag{1.9}$$

$$u^{(n)} = L_p^{(0)} A^{(n)} \widetilde{L}_p^{(0)}. \tag{1.10}$$

$$\Delta u_{kl}^{(n)} = Sp\left[\Delta P^{(n)} \frac{\partial^2 H_e}{\partial q_k^{(0)} \partial q_l^{(0)}}\right] + \frac{1}{2} Sp\left[\frac{\partial \Delta P^{(n)}}{\partial q_k^{(0)}} \frac{\partial H_e}{\partial q_l^{(0)}} + \frac{\partial \Delta P^{(n)}}{\partial q_l^{(0)}} \frac{\partial H_e}{\partial q_k^{(0)}}\right]. \tag{1.11}$$

Assuming that deformations of molecular model associated with electronic excitation are small (compared to equilibrium configuration) and retaining only the first order terms in (1.7), we then replace $A^{(n)}$ by $A^{(0)}$ getting rid of the term proportional to $\Delta A^{(n)} \Delta a^{(n)}$. This means that for calculation of changes in geometry the matrices of force constants and squared frequencies are taken equal to those of the ground state (in doing so, the errors in $\Delta s^{(n)}$ do not exceed ~10%). Expressions (1.2) and (1.8) give:

$$\Delta s^{(n)} = -L_q^{(0)} \Lambda^{-1} \Delta a^{(n)} = -L_q^{(0)} \Lambda^{-1} \widetilde{L}_q^{(0)} Sp\left[\Delta P^{(n)} \frac{\partial H_e}{\partial q^{(0)}}\right]. \tag{1.12}$$

Similarly, neglecting the derivatives of density matrix we write:

$$\Delta u_{kl}^{(n)} = Sp\left[\Delta P^{(n)} \frac{\partial^2 H_e}{\partial q_k^{(0)} \partial q_l^{(0)}}\right]. \tag{1.13}$$

These approximations are also supported by the fact that the relation between changes in density matrix, geometry and force field parameters have linear form, which is well consistent with correlations between bond lengths, force constants and bond orders (see, for example, [14, 16, 21]).

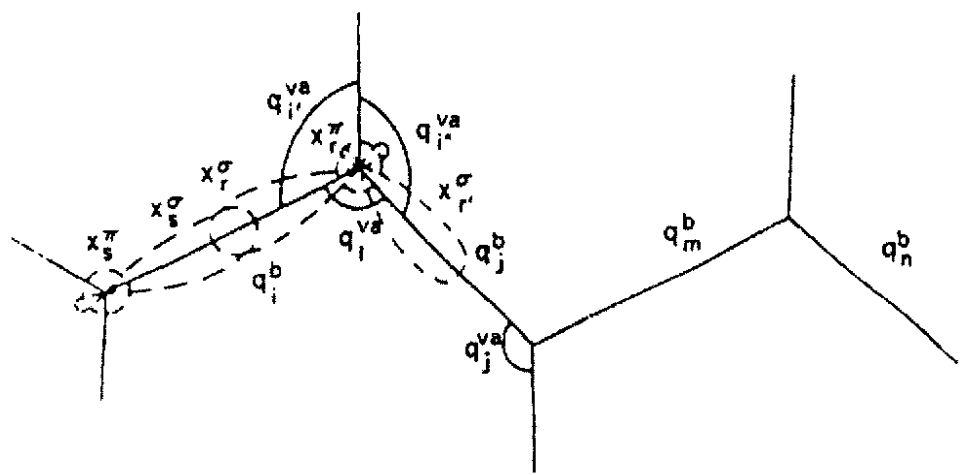

Fig.1.1. HAOs and internal coordinates of molecular fragment.

Thereby, the derivatives $\partial H_e/\partial q_k$, $\partial^2 H_e/\partial q_k \partial q_l$ taken at equilibrium configuration define the parameters of adiabatic molecular model and serve as "hidden" parameters of the theory. Analysis of their properties at the qualitative level [58] had shown that this parameterization provides good starting point for the semiempirical theory of vibronic spectra based on the use of molecular fragments. Here we discuss the most important features of the proposed method and models.

The locality of parameters is determined by the local properties of Coulomb and resonance integrals that come from many quantum calculations. Since matrix elements $H_{rs} \equiv \langle \chi_r | H_e | \chi_s \rangle$ (where $\chi_r$ represents HAO basis) decrease rapidly as the distance from atoms $\alpha_r$ and $\alpha_s$, on which HAO $\chi_r$ and $\chi_s$ are centered, increases, these parameters can be attributed to local groups of atoms. Moreover, differentiation with respect to natural coordinates enhances this effect, so that only interactions between HAO and atoms, which form corresponding natural coordinate, will contribute to the value of parameter. In fact, this important property allows to assign parameter to a certain small (3–4 atoms) molecular fragment. The set of quantities $\partial H/\partial q$ can then be divided into groups depending on the type of HAO ($\sigma$ or $\pi$) and natural coordinate (bond length, valence angle, etc.). For example, $\partial H_{rs}^t / \partial q_i^c$, where $t$ refers to the type of HAO, $c$ denotes the type of natural coordinate, $r$ and $s$ are the indices of HAO, $i$ is the number of natural coordinate (see Fig.1.1).

Besides, these properties make it possible to rank quantities $\partial H_{rs}/\partial q_i$, $\partial^2 H_{rs}/\partial q_i \partial q_j$ and take into account only major parameters. For example, estimates give:

$$\left|\frac{\partial H_{rs}^{\sigma,\pi}}{\partial q_i^b}\right| \gg \left|\frac{\partial H_{rr}^{\sigma}}{\partial q_i^b}\right| > \left|\frac{\partial H_{rr}^{\sigma}}{\partial q_j^b}\right| \approx \left|\frac{\partial H_{rr}^{\sigma}}{\partial q_i^a}\right| > \left|\frac{\partial H_{rr}^{\pi}}{\partial q_{i,i',i''}^a}\right| \approx \left|\frac{\partial H_{rs}^{\sigma}}{\partial q_i^a}\right| \gg \left|\frac{\partial H_{rr}^{\pi}}{\partial q_{i,j}^b}\right| > ...,\qquad(1.14)$$

$$\left|\frac{\partial^2 H_{rs}^{\sigma,\pi}}{\partial q_i^{b\,2}}\right| \gg \left|\frac{\partial^2 H_{rr}^{\sigma}}{\partial q_i^{b\,2}}\right| \gg \left|\frac{\partial^2 H_{rr}^{\sigma}}{\partial q_j^{b\,2}}\right| \approx \left|\frac{\partial^2 H_{rr}^{\pi}}{\partial q_{i,j}^{b\,2}}\right| \approx \left|\frac{\partial^2 H_{rs}^{\sigma,\pi}}{\partial q_{i,k}^{a\,2}}\right| \approx \left|\frac{\partial^2 H_{rr}^{\sigma}}{\partial q_i^{a\,2}}\right| \gg \left|\frac{\partial^2 H_{rr'}^{\sigma}}{\partial q_i^{a\,2}}\right| \approx \left|\frac{\partial^2 H_{rr'}^{\sigma}}{\partial q_{i,j}^{b\,2}}\right| > ...\;.\qquad(1.15)$$

Note also that more accurate ranking by *ab initio* calculations does not usually make much sense, cause optimal values can be obtained only through semiempirical calibration with real experimental spectroscopic data.

It is possible, therefore, to work out parametric models of different levels of accuracy considering only the terms up to a certain order that significantly reduces the number of parameters responsible for vibronic band shape and brings it to a small number. This not only simplifies calculations of (1.12) and (1.13), but also makes the model intuitive and assures that determination of parameters from experimental data represents a well-posed problem.

Yet another important property of the proposed system of parameters is that it is unified for different electronic states. All the parameters are defined by the molecular structure and can be obtained from purely geometric considerations, whereas the changes occurring due to electronic excitation are accounted by the increment in electron density matrix $\Delta P^{(n)}$ which can be calculated by one of quantum chemical methods.

Further, transferability of $\partial H_{rs}/\partial q_i$, $\partial^2 H_{rs}/\partial q_i \partial q_j$ in homologous series follows from the same property of Coulomb and resonance integrals. The derivatives are not small, so that the requirements on their transferability are generally weaker than that for force field and geometrical parameters of PES. These requirements are basically of the same order as for force constants in the theory of IR spectra, and, hence, transferability of the parameters is expected to be even better (it is found sufficient for quantitative predictions of vibronic spectra), cause these quantities, unlike force constants, do not depend upon the effect of far interactions, e.g., in conjugated systems.

It is clear enough that matrix $H_e$ should be computed in HAO, but not in AO, basis, since in this case structural features of molecular fragments appear explicitly. It also makes the system of parameters invariant and improves its transferability by eliminating the dependence on the choice of laboratory coordinate system.

The parameters have clear physical meaning and can be computed by *ab initio* or semiempirical methods. For example, $\partial H_{rs}^{\pi}/\partial q_i^b$ describes the change in energy of $\pi$-electron distributed along the bond $q_i^b$ according to $(\chi_r \chi_s)$ upon variation of this bond, while $\partial H_{rr}^{\sigma}/\partial q_i^b$ is related to energy of interaction of the electron distribution $(\chi_r^{\;2})$ with the atom $\alpha_s$ being a part of the bond $q_i^b$. In contrast to empirical scaling factors used in some quantum chemical methods [59], the values of these parameters can be preliminary estimated by direct quantum calculations. However, there is no reason for semiempirical parameters to coincide with those calculated *ab initio*, mainly because of approximations involved from both sides.

Calculations done for specific molecular models (butadiene, hexatriene, octatetraene), all having characteristic periodic fragment $\overset{H}{>}C=$, have confirmed qualitative *a priori* estimates [48]. The parameters demonstrate pronounced local properties; those directly associated with $\overset{H}{>}C=$ group clearly dominate over all the rest which are smaller by an order of magnitude or more (for hexatriene $\partial H_{rs}^\pi/\partial q_i^b = 3(\partial H_{rs}^\pi/\partial q_j^b) = 7(\partial H_{rs}^\pi/\partial q_m^b) = 20(\partial H_{rs}^\pi/\partial q_n^b)$) preserving their values within molecule and across homologous series with maximal deviation of 50%. We can then reduce the number of parameters by retaining leading quantities and neglecting small ones, as in the following sequence (given in parentheses are averaged values): $\partial H_{rr}^\sigma/\partial q_j^b$ (2 a.u.), $\partial H_{rr}^\pi/\partial q_{i,j}^b$ (0.8 a.u.), $\partial H_{rr}^\sigma/\partial q_i^b$ (0.5 a.u.), $\partial H_{rs}^\pi/\partial q_i^b$ (0.3 a.u.), $\partial H_{rr}^\pi/\partial q_i^a$ (0.3 a.u.), $\partial H_{rr}^\pi/\partial q_{i',i''}^a$ (-0.15 a.u.). This allows to introduce $\overset{H}{>}C=$ fragment along with its intrinsic small system of parameters as a transferable building block for models of more complex molecules containing polyene fragments.

It is essential that $\Delta s$, $\Delta u$ (1.12), (1.13) depend not only on values of these parameters, but also on changes in density matrix $\Delta P^{(n)}$. Its typical structure for $\pi\pi^*$ transitions in conjugated polyatomic molecules (as in polyenes, for example) suggests that principle contributions (three orders of magnitude higher than others) to $\Delta s$, $\Delta u$ are specifically attributed to $\partial H_{rs}/\partial q_i^b$ (related to molecular geometry) and $\partial^2 H_{rs}/\partial q_i^{b2}$ (responsible for changes in force field upon electronic excitation). So, to a first approximation, these two quantities define the model of excited electronic states of $\pi$-electron molecular systems, which correlates well with direct calculations of vibronic spectra of polyenes, acenes, diphenylpolyenes and related species with $\pi\pi^*$ transitions [21-24]. The results of these simulations, which take into account only changes in bond lengths and force constants, are in good agreement with experiment. However, it is expected that quantitative description of $n\pi^*$ transitions accompanied by substantial deformation of $\sigma$-electron density will require full consideration of changes in valence angles and corresponding parameters.

For a parametric method and computational algorithms to be applicable in practice it is required to satisfy some stability criteria with respect to variations of parameters. It can be shown that stability (in the sense that small variations of model do not lead to large changes in spectral curve) is guaranteed by the stability with respect to "natural" PES parameters $\Delta s$, $\Delta u$ and linearity of relations (1.12), (1.13).

Thus, here we see that, due to the reasons discussed, straightforward conventional adiabatic molecular models built in terms of bond lengths, valence angles, force constants proved inappropriate for calculations of vibronic spectra. In turn, the new system of "hidden" parameters has been designed to meet the requirements that appear in practical modeling and direct simulations. This system itself holds much promise as a basis for parametric semiempirical adiabatic theory of vibronic spectra and efficient methods for computing vibronic transition probabilities. Certainly, the approach should be tested through extensive comparative studies over a number of molecules and homologous series. The next section outlines our results in this direction. We examine properties and efficiency of the proposed parametric method and its predictive ability making comparisons of calculated vibronic spectra with experimental data.

## 2. Calculations of stationary vibronic spectra with the parametric method

Of the homologous series that demonstrate well resolved vibrational structure of electronic spectra, extensively studied, therefore, both theoretically and experimentally, are polyenes, their substituted derivatives (in particular, diphenylpolyenes), acenes and azines. The results collected here show the parametric method as it was employed for calculations of vibronic spectra, specifically for polyenes and acenes.

### 2a. Polyenes

Molecular models of polyenes in their ground states were constructed from fragments of smaller homologous with force field parameters transferred "as is". It was confirmed by the simulations that, in an analogy with polymers, as the length of polyene chain increases, contributions from terminal groups gradually become negligible, whereas parameters of the inner $\overset{H}{>}C=$ fragments approach a limit, which can be considered constant for all groups within a molecule and for longer homologues. For this reason, values of bond lengths and force constants for links between fragments were set to those of inner groups in sufficiently long molecules. Experimental data, other theoretical studies including quantum calculations of electron density and estimates for parameters obtained through correlations "index—length—force constant of the bond" [21] have justified this procedure. Calculations of vibrational frequencies of the ground state have shown that this simple approximation provides good agreement with experiment, so that there is no need for solving inverse problems. This supports, again, the fact that the idea of molecular fragments together with understanding the peculiar properties of various homologous series give efficient models of the ground states with accurately computed IR frequencies.

By virtue of approximations, any parametric method is a semiempirical one that implies "adjustment" of the system of parameters to match direct or indirect experimental data. Such corrections have been introduced to rough values of $^H\!\!>\!C\!=$ parameters obtained by CNDO/S method in order to account for real errors associated with approximations, inaccuracies of quantum calculations and adiabatic model itself. In the first approximation of the theory (with a minimal set of parameters), this gives $\partial H_{rs}^{\pi}/\partial q_i^b$ =0.055 a.u., $\partial^2 H_{rs}^{\pi}/\partial q_i^{b^2}$ =0.1 a.u. Making the use of a single parameter, which was set equal for all groups and all homologous ($N_{CC}$=3–13, [48, 52, 53]), allows, however, to achieve satisfactory quantitative description of changes in electronic states upon excitation. For bond lengths the deviations are ~15%, in average, with a maximum of ~35% for C–C bond in hexatriene (variations of the same order was observed in the solutions of inverse problems [25]). The same holds for force constants. It is important that the parametric approach treats changes of different natural coordinates, including angles, in a unified way. These changes for valence angles come out much smaller than those for bond lengths (about 20%, in relative units), which is consistent with typical results of inverse problem solution (15% [25]). Moreover, for the majority (~70%) of angular coordinates, the method in most cases reproduces these results quantitatively.

Primary criteria for correctness of parametric approach, chosen system of parameters and specific values of parameters, adequacy of excited state model, should necessarily be based on agreement of calculated spectra with experiment. This is because there are no direct experimental methods that would measure PES parameters and the only indirect way to get this information is spectroscopic interpretation. The other part of the reason is that computational uncertainties of quantum chemical methods turn out to be of the order of changes in these quantities upon electronic excitation and so there is just no point to compare such results obtained, for example, by different methods.

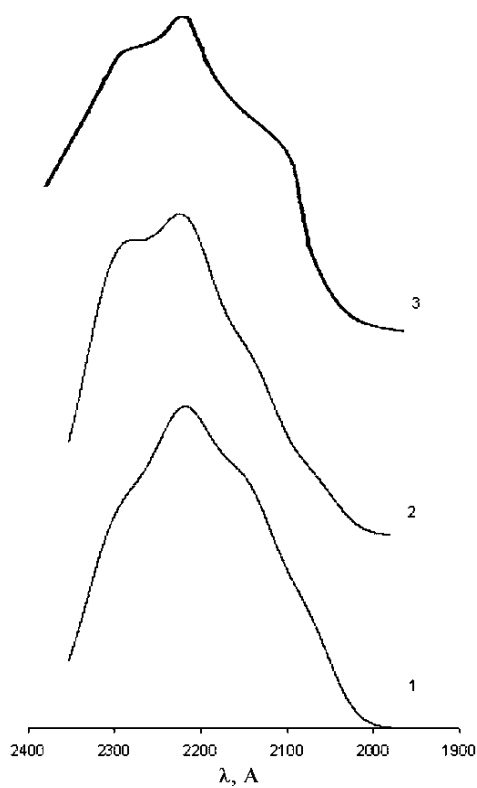

Fig.2.1. Absorption spectra of butadiene calculated in the first (1) and second (2) approximations and experimental absorption spectrum (3) [60].

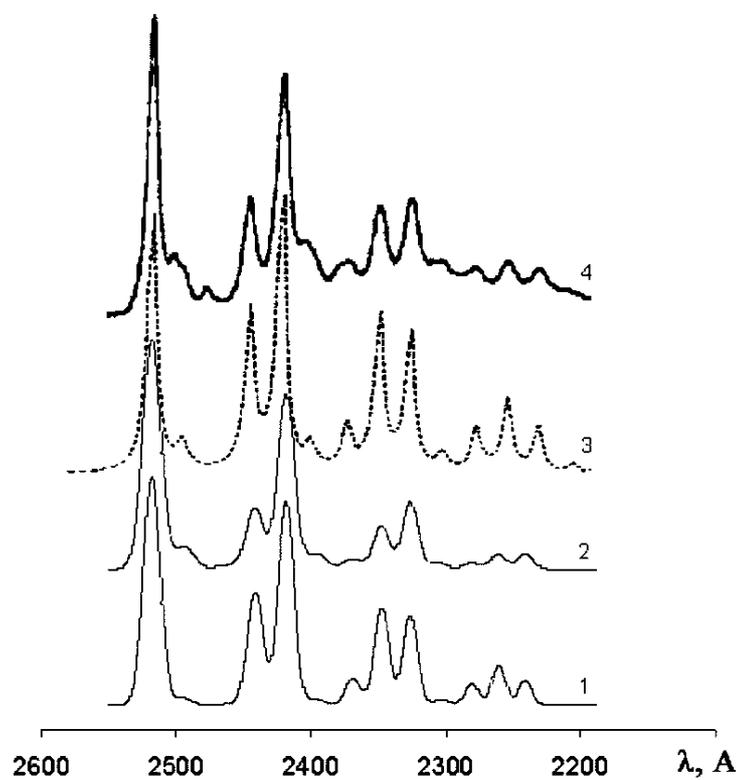

Fig.2.2. Absorption spectra of hexatriene calculated in the first (1) and second (2) approximations; using the PPP–CI model (3); and experimental absorption spectrum (4) [61].

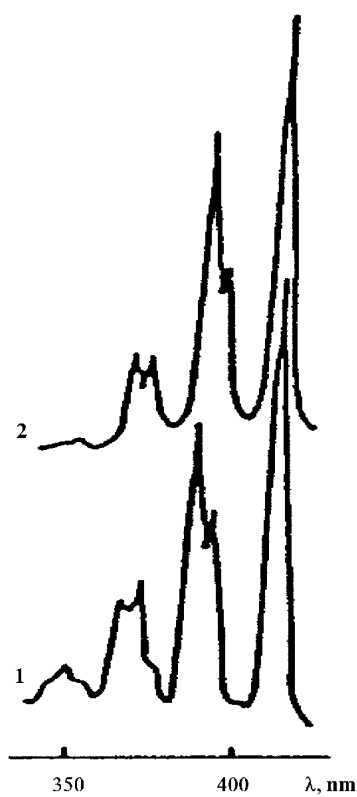

Fig.2.3. Experimental (1) [62] and calculated in the first approximation (2) absorption spectra of hexadecaheptaene.

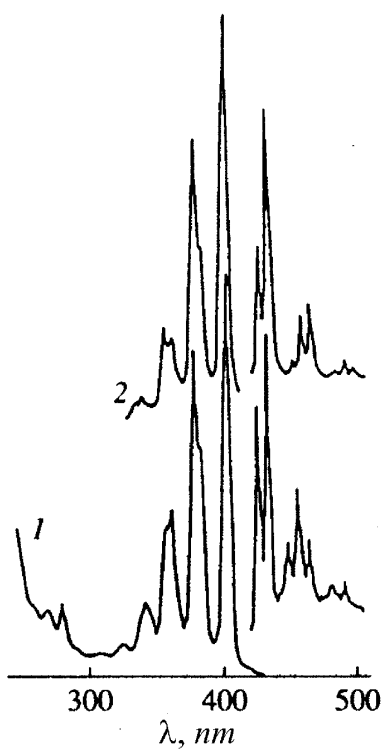

Fig.2.4. Experimental (1) [62] and calculated in the first approximation (2) absorption and fluorescence spectra of tetradecaheptaene.

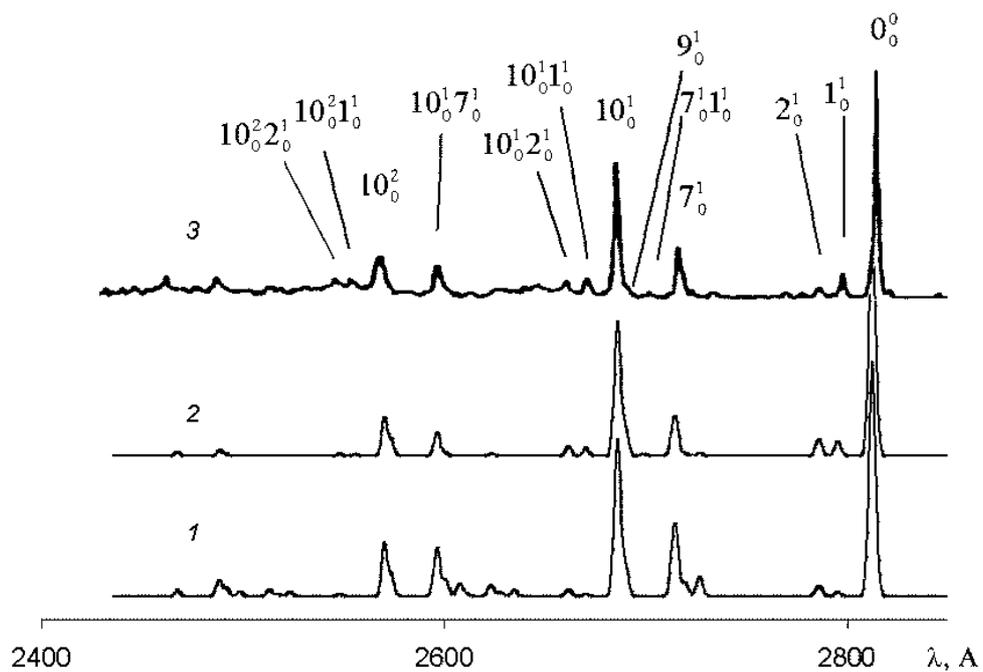

Fig.2.5. Absorption spectra of octatetraene calculated in the first (1) and second (2) approximations [52] and experimental absorption spectrum (3) [63].

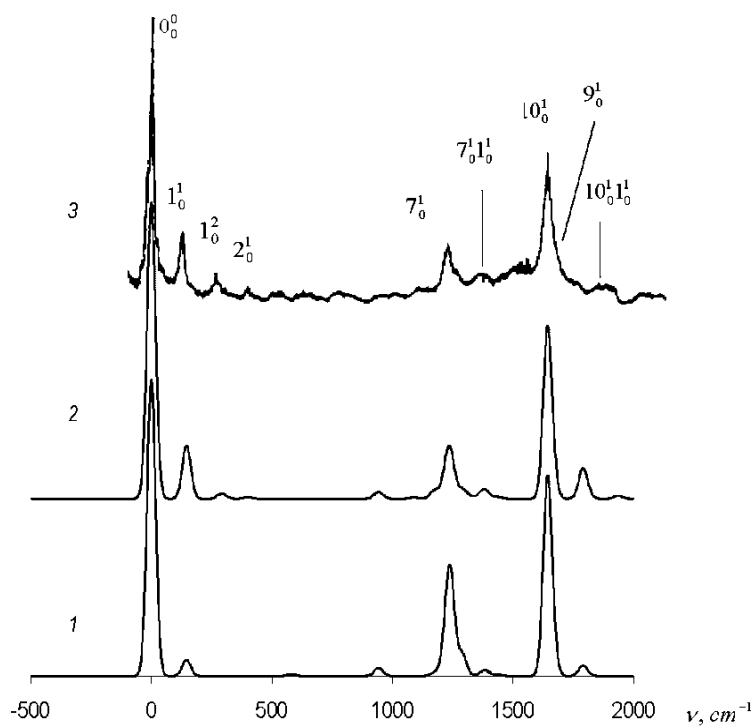

Fig.2.6. Absorption spectra of decatetraene calculated in the first (1) and second (2) approximations and experimental absorption spectrum (3) [64].

Calculated absorption spectra of polyenes (butadiene, hexatriene, hexadecaheptaene, tetradecaheptaene, octeraene, and decatetraene) shown in Fig.2.1-2.6 demonstrate remarkable quantitative agreement in band positions and intensities with experimental spectra. Note that the first approximation model was calibrated for butadiene and hexatriene and then used unchanged. Therefore, calculations of spectra of longer polyenes in this approximation are purely predictive.

There are some peculiarities, though, to explain which we should go beyond the first approximation. For low-frequency bands calculations give intensities systematically lower than observed. For example, for octatetraene experimental relative intensity at ~200 cm$^{-1}$ is about 0.11, whereas computed value is as low as 0.02; for hexatriene at ~300 cm$^{-1}$ these are 0.18 and 0.03, respectively. On the contrary, C-C bond stretching vibronic modes (~1200 cm$^{-1}$) in calculated spectra of octatetraene and butadiene are perceptibly stronger than in experimental (0.41 and 0.23 for octatertaene). This leads to significant errors in intensities of overtones and combination bands (for example, the first overtone of 1235 cm$^{-1}$ vibration in octatetraene spectrum is about fourfold more intense than its experimental value). It was found reasonable to assume that these discrepancies are attributed to inaccurate geometries of excited state models, namely to changes in angular coordinates, since their contributions even to valence modes may be noticeable, ~10%. Similar result follows also from the analysis performed in [48], where it was shown that consecutive refinement of first approximation model (by first taking into account changes in bond lengths only and, then, in angles) improves agreement between theoretical and experimental spectra.

Of the short polyenes (butadiene, hexartiene, octatetraene), spectrum with best resolved vibrational structure was obtained experimentally for octatetraene [63]. Besides, it can be expected that with increasing length of polyene chain, the number of relevant angular parameters will grow and that overall spectral effect for octatetraene will be stronger, than it is for shorter homologues. For these reasons, we calibrated parameters of the second approximation using octatetraene model as a testing bed, with general strategy being to apply unchanged set of parameters to other polyenes. This allowed us to test the quality of parameterization, transferability of parameters and predictive ability of the method itself.

The best results in the second approximation was obtained with the following: $\partial H_{rs}^{\pi}/\partial q_i^{b} = 0.045\,\text{a.u.}$ ; $\partial H_{rr}^{\pi}/\partial q_i^{a_{CCC}} = -0.005\,\text{a.u.}$ ; $\partial H_{rr}^{\pi}/\partial q_i^{a_{CCH}} = -0.0015\,\text{a.u.}$ ; $\partial^2 H_{rs}^{\pi}/\partial q_i^{b2} = 0.1\,\text{a.u.}$. As this sequence indicates, the first approximation parameter $\partial H_{rs}^{\pi}/\partial q_i^{cB}$ undergoes only small change (<20%), while the angular parameters are by an order of magnitude smaller, that justifies our preliminary estimates, ranking and possibility of parametric theories of different levels of accuracy [48].

Calculated spectral curve, frequencies, intensities and line assignment in vibronic absorption spectrum are given on Fig.2.5 (curve 2). Theoretical spectrum is in good agreement with intensities of all major peaks observed in the experiment. In the second approximation, the intensities of $10_0^1$ and $7_0^1$ bands come out smaller approaching experimental values. Further, equal attention should be paid to weak lines; as compared to the first approximation, the sum of relative discrepancies between experimental and calculated intensities reduces from 0.7 down to 0.4, their mean value decreases from 0.05 to 0.03, standard deviation — from 0.05 to 0.02. Thereby, the second approximation evidently gives better results and describes more accurately the structure of excited states and, especially, its features related to angular deformations.

We then applied the second approximation model, without making any corrections, to other homologous — butadiene (Fig.2.1), hexatriene (Fig.2.2) and decatetraene (Fig.2.6). For all the molecules, it improves agreement with experiment as, for example, in low-frequency (~300 ÷ 400 cm$^{-1}$) range of hexatriene spectrum, as well as around major vibronic lines (~1200 cm$^{-1}$ and ~1600 cm$^{-1}$) and for combination bands (compare experimental (4), first (1) and second (2) approximation curves on Fig.2.2). The effect is also noticeable in poorly resolved absorption spectrum of butadiene (Fig.2.1).

Highly structured spectrum of decatetraene [64] (Fig.2.6) is shaped mainly by characteristic peaks $1_0^1$, $2_0^1$, $7_0^1$ and $10_0^1$ corresponding to excited state vibrations with frequencies $\nu_e$=132, 405, 1230 and 1651 cm$^{-1}$ (for mode numbers refer to [52]). Calculated values of these frequencies are $\nu_c$ =146, 403, 1239 and 1641 cm$^{-1}$, respectively, that shows good quantitative match with experimental picture. Weak bands $5_0^1$, $6_0^1$ and $8_0^1$ overlap with stronger line $7_0^1$ giving rise to complex structure in this region, well reproduced in simulations. While modes 1 and 2 represent deformation vibrations, modes 7 and 10 contain a mix of CC bond stretchings and CCC and CCH angular vibrations. So, it comes as no surprise that in the first approximation with complete neglect of angular parameters, computed intensities differ markedly from experimental ones ($I_e(7_0^1)$=0.2 and $I_c(7_0^1)$=0.32, $I_e(10_0^1)$=0.52 and $I_c(10_0^1)$=0.63, $I_e(1_0^1)$=0.2 and $I_c(1_0^1)$=0.05, $2_0^1$ vibration is not present in the spectrum at all). Note how going to the second approximation improves these things: $I_c(7_0^1)$=0.14, $I_c(10_0^1)$=0.54, $I_c(1_0^1)$=0.18, also progression $1_0^2$ and low-frequency band $2_0^1$ appear; the sum of relative differences between theoretical and experimental intensities reduces fourfold (from 0.38 to 0.1). Characteristic shape of $7_0^1$ band approaches experimental one and can now be explained by the presence of weak, unresolved in experiment, adjacent lines $I_c(1260$ cm$^{-1})\cong 0.01$ and $I_c(1289$ cm$^{-1})\cong 0.03$ that correspond to totally symmetric vibrations. Similarly, the origin of $10_0^1$ asymmetry can be traced to existence of $9_0^1$ satellite with $I_c(1662$ cm$^{-1}) \cong 0.1$. In our analysis, as distinct from [64], 273 cm$^{-1}$ band is assigned to the second harmonics $1_0^2$.

Yet another evidence for the parametric method to be efficient at describing fine spectral effects in large molecules (some tens of atoms) comes from consideration of methyl-substituted polyenes. Indeed, detected spectral shifts associated with methyl-substitution match experimental results within accuracy less than 3% and correctly reproduced shift directions. For example see octatetraene (OT), Fig.2.5, and decatetraene (DT), Fig.2.6: the strongest $10_0^1$ band ($\nu_e^{OT}=1645$ cm$^{-1}$), which is almost unaffected, shifts so that $\nu_c^{OT}/\nu_c^{DT}=1.02$ versus $\nu_e^{OT}/\nu_e^{DT}=1.0$, while low-frequency deformation vibrations $1_0^1$ ($\nu_e^{OT}=197$ cm$^{-1}$) and $2_0^1$ ($\nu_e^{OT}=348$ cm$^{-1}$), sensitive to substitution, show $\nu_c^{OT}/\nu_c^{DT}=1.5$ versus $\nu_e^{OT}/\nu_e^{DT}=1.49$ and $\nu_c^{OT}/\nu_c^{DT}=0.85$ versus $\nu_e^{OT}/\nu_e^{DT}=0.86$, respectively. Similar trends were observed for intensities. Both experimental and calculated intensities at $10_0^1$ are ~10% smaller for decatetraene ($I_c^{OT}=0.57$, $I_c^{DT}=0.54$ and $I_e^{OT}=0.6$, $I_e^{DT}=0.52$); $9_0^1$ band ($\nu_c^{OT}=1622$ cm$^{-1}$, $I_c^{OT}=0.11$ and $\nu_c^{DT}=1662$ cm$^{-1}$, $I_c^{DT}=0.08$) undergoes a shift by ~40 cm$^{-1}$ upon substitution and appears as either low-energy (OT) or high-energy (DT) shoulder near $10_0^1$, reproducing correctly the observed line shape in this region. The next strong composite profile around $7_0^1$ changes mainly in intensity ($\Delta\nu \leq 3\%$, $\Delta I \approx 15\%$), due to the growth of weaker overlapping $6_0^1$ band. Low-frequency $1_0^1$ component exhibits the maximal change in intensity (2 times), as well as in line position, upon methyl-substitution.

Simulations have shown that as the length of polyene chain increases, general spectral pattern stays the same, but some redistribution of intensity between major bands occurs, so that intensities of valence modes (~1650 and ~1250 cm$^{-1}$) scale approximately as $I = 1.56 N_{CC}^{-0.5}$ (accurate within $\Delta I < 5\%$), while intensities of deformation modes ($\leq 500$ cm$^{-1}$) as $I = 0.003 N_{CC}^{1.7}$ ($\Delta I < 25\%$), where $N_{CC}$ is the number of C–C bonds. This is an indication of the importance of angular parameters (second approximation) for simulations of longer polyenes.

Here we see that parametric method, even with minimal number of parameters (only 2 in the first approximation), allows to develop quantitative transferable models of polyenes in excited states, perform spectroscopically accurate predictive calculations and detailed interpretation of spectra. An advantage of the parametric method at capturing major features of excited states is illustrated on Fig.2.2, where curve 3 shows the same hexatriene spectrum simulated by alternative quantum chemical method [63, 65, 66]. Similar results confirmed this for other molecules too [49, 50].

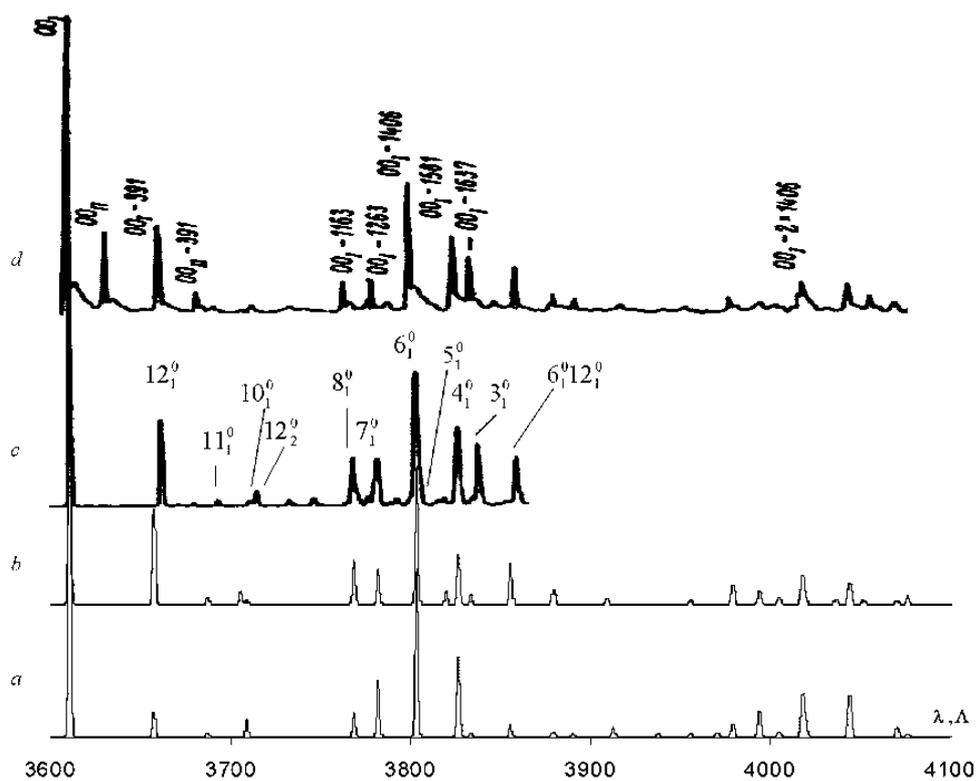

Fig.2.7. Fluorescence spectra of anthracene calculated in the first (a) and second (b) approximations and experimental fluorescence spectra in supersonic jet at 00-excitation (c) [67] and 4K *n*-hexane (d) [68].

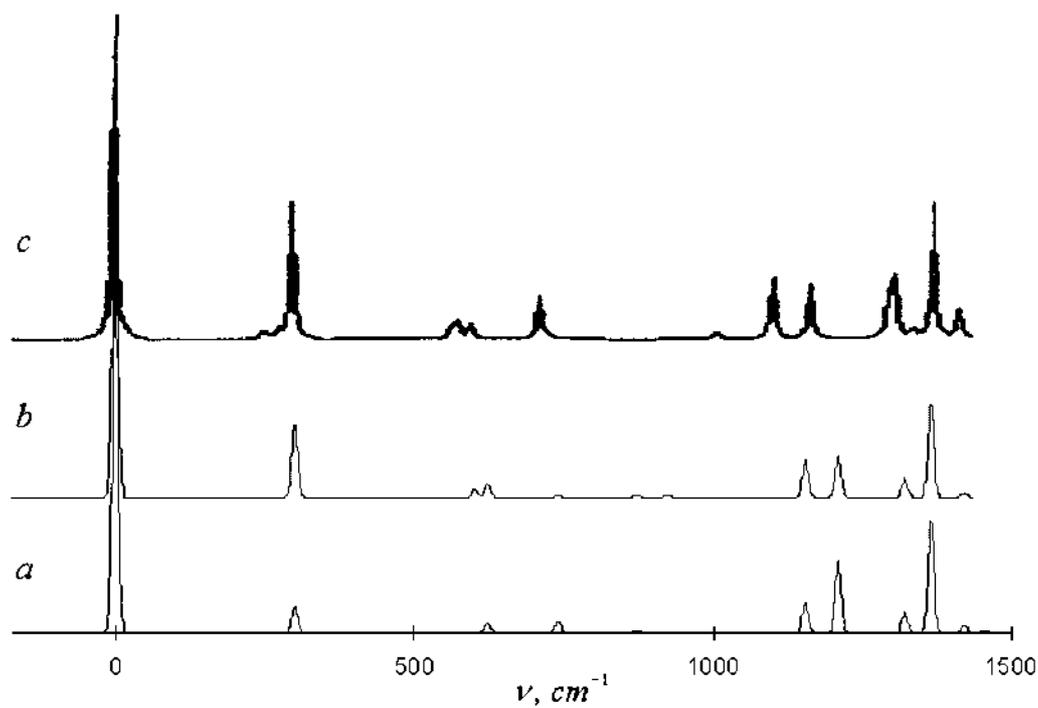

Fig.2.8. Absorption spectra of tetracene calculated in the first (a) and second (b) approximations and experimental absorption spectrum (c) [69].

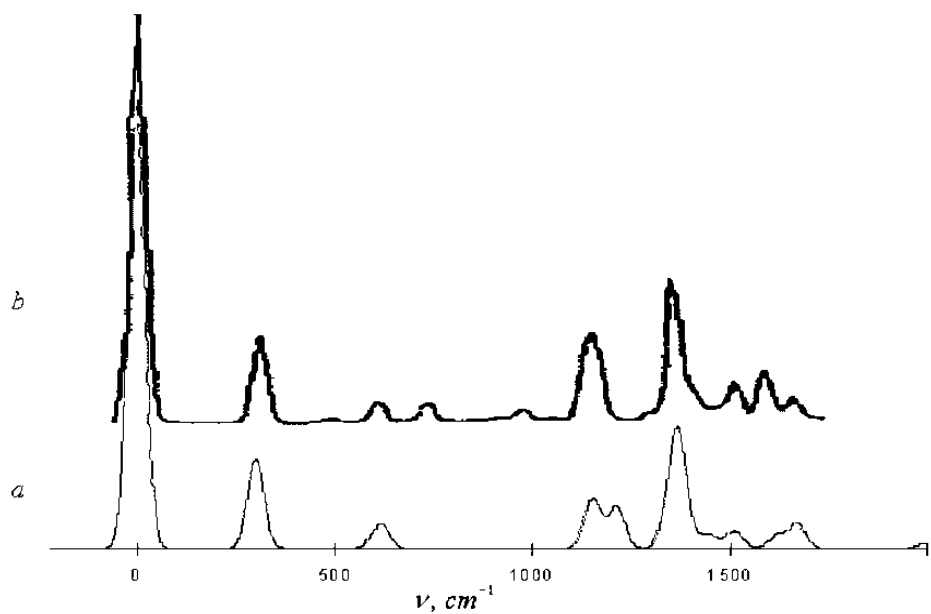

Fig.2.9. Calculated in the second approximation (a) and experimental (b) [68] fluorescence spectra of tetracene.

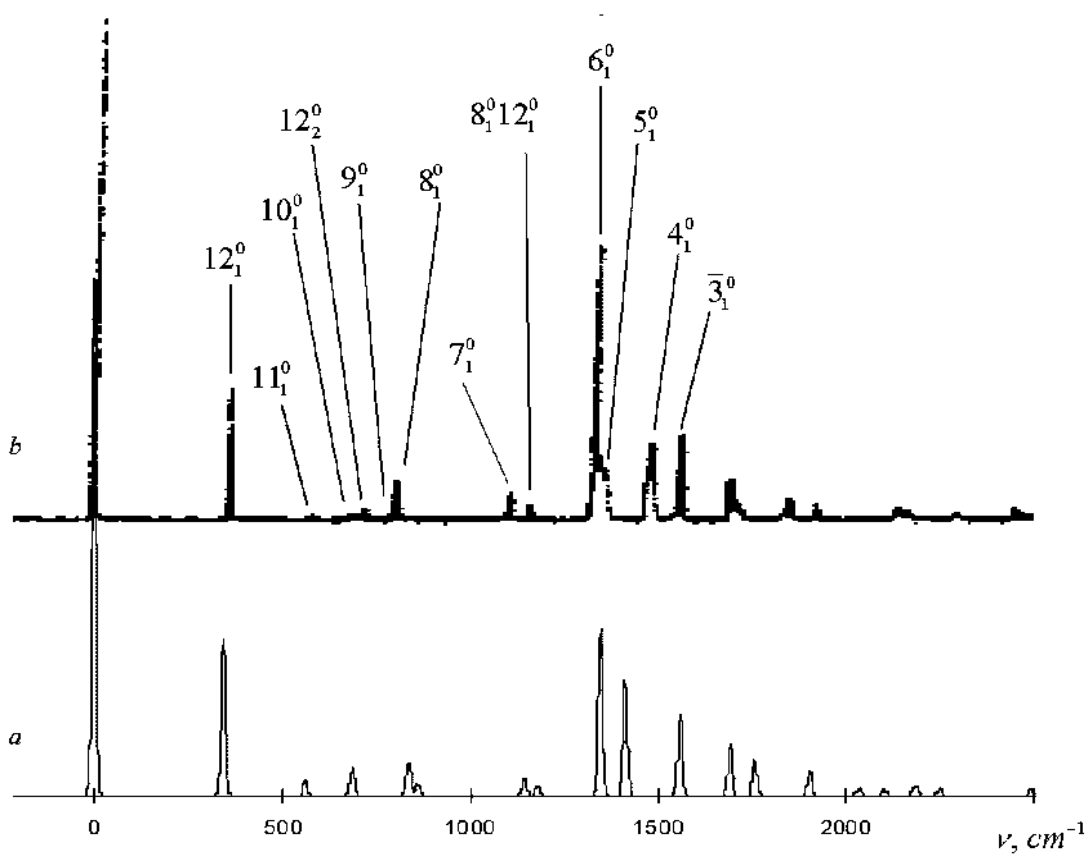

Fig.2.10. Calculated in the second approximation (a) and experimental in supersonic jet upon 00-excitation (b) [67] fluorescence spectra of anthracene-$d_{10}$.

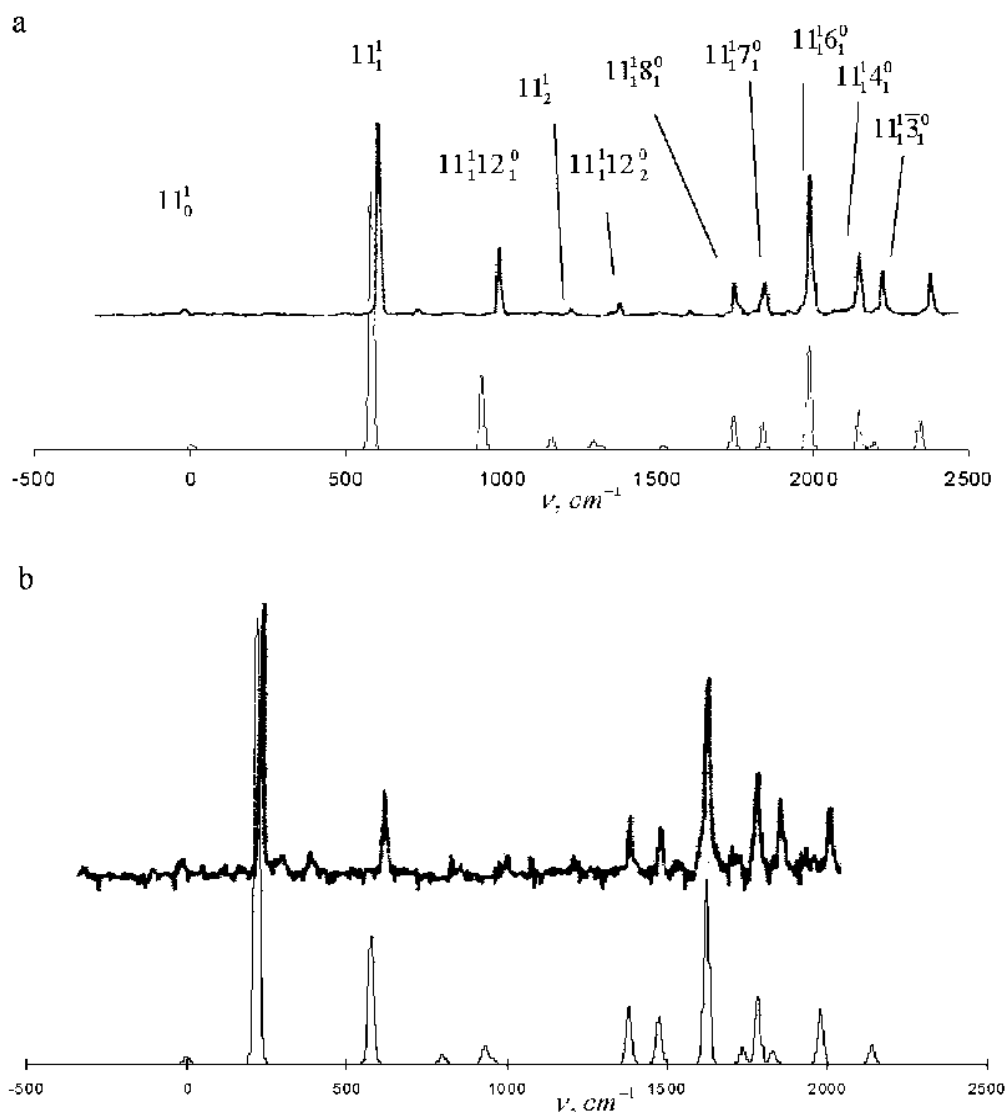

Fig.2.11. Calculated (lower) and experimental (upper) [67] dispersed fluorescence spectra of anthracene at $S_1+583$ cm$^{-1}$ (a) and $S_1+237$ cm$^{-1}$ (b) excitation.

## 2b. Acenes

Comparison of calculated in the first approximation [49] spectrum of anthracene (Fig.2.7a) with the results of experiments (Fig.2.7c,d) [67, 68] finds good overall agreement up to weak totally symmetric modes. However, we note significant differences, for example, in intensities of $12_1^0$ ($\nu_e$=391 cm$^{-1}$), $8_1^0$ ($\nu_e$=1163 cm$^{-1}$) and $6_1^0 12_1^0$ ($\nu_e$=1797 cm$^{-1}$). The sum of discrepancies reaches 0.48 in relative units. For tetracene, essentially the same picture is observed (Fig.2.8a,c).

It seemed *a priori* reasonable that for polycyclic compounds, in which changes of C–C bonds are often accompanied by angular deformations, the second approximation with its angular parameters (first derivatives with respect to CCC and CCH angles) will be expected to improve accuracy. The system of parameters was optimized for anthracene (see the best fit obtained, Fig.2.7a,b) and basically takes the form of small (~10%) corrections to the first approximation model: $\partial H_{rs}^\pi / \partial q_i^{b_1} = 0.07$ a.u., $\partial H_{rr}^\pi / \partial q_i^{a\,CCC} = 0.008$ a.u., $\partial H_{rr}^\pi / \partial q_i^{a\,CCH} = -0.002$ a.u. and $\partial^2 H_{rs}^\pi / \partial q_i^{b_2} = 0.3$ a.u.. As Fig.2.7b indicates, corrected approximation reproduces all vibronic bands present in the experimental spectrum, except for those (for example, $3_1^0$ at $v_e$=1643 cm$^{-1}$) that was previously interpreted as related to either nontotally symmetric $b_{1g}$ vibrations [67] or different matrix site [68]. This is consistent with Franck-Condon approximation and confirms correct assignment of these bands. Detailed comparison has shown that going to the second approximation reduces overall sum of relative differences between calculations and experiment from 0.48 до 0.13, mainly due to better agreement in $12_1^0$ ($v_e$=391 cm$^{-1}$) and $6_1^0$ ($v_e$=1406 cm$^{-1}$) bands, deformation modes $7_1^0$ ($v_e$=1263 cm$^{-1}$) and $8_1^0$ ($v_e$=1163 cm$^{-1}$), second harmonics around 2500–3500 cm$^{-1}$. In particular, the second harmonics of the 7th and 8th vibrations appear symmetrically about $4_1^0$ in the calculated spectrum (Fig.2.7b), but in the experimental spectrum they are masked by stronger $4_1^0$ and $3_1^0$ bands.

Following the same strategy, we tested efficiency and predictive ability of the system of parameters by transferring it to anthracene-d$_{10}$ and tetracene (Fig.2.8-2.10) with no modifications introduced.

Computed shifts in vibrational frequencies of anthracene upon deuteration are typically less than 3%, which is fairly close to the observed effect. Among active vibronic modes in the range 1000–1300 cm$^{-1}$, the largest shifts (100–320 cm$^{-1}$) are detected for those with significant contributions from CCH angular vibrations. The intensities are also affected by a factor ranging from 2 to 6 (see Fig.2.7 and Fig.2.10); $8_1^0$ ($v_e$=1163 cm$^{-1}$) demonstrates $\Delta v_c = 325\,cm^{-1}$ ($\Delta v_e = 315\,cm^{-1}$) and $\Delta I_c = -0.07$ ($\Delta I_e = -0.1$), $7_1^0$ ($v_e$=1263 cm$^{-1}$) changes by $\Delta v \approx -100\,cm^{-1}$ with intensity decrement $\Delta I_c = -0.08$ ($\Delta I_e = -0.07$). Conversely, intensity of $5_1^0$ band of anthracene-d$_{10}$ spectrum is larger both in experiment (0.13) and calculations (0.23). Frequency shifts for other vibrations, including those that correspond to the strongest components in anthracene spectrum (for instance, $12_1^0$ at $v_e$=391 cm$^{-1}$), do not exceed 50 cm$^{-1}$. Their intensities are found almost invariant to deuteration. These factors preserve characteristic structure of anthracene spectrum. Note good overall accuracy achieved in simulations of spectral effects caused by deuteration that validates also basic model of anthracene excited states in the second approximation.

Shown on Fig.2.8,2.9 are simulated spectra of tetracene. Both curves (first and second approximations, Fig.2.8) fit well experimental data, but evidently the second approximation provides closer match. See low-frequency band $v_e$=314 cm$^{-1}$, complex structure around 600 см$^{-1}$, two satellites in the region 1000–1200 cm$^{-1}$. Half-width of ~1200 cm$^{-1}$ band is larger in the experimental spectrum than the others, that suggests existence of two adjacent lines with comparable intensities revealed by simulations. Note especially that good agreement with the experiment for tetracene was obtained with the system of parameters taken "as is", i.e. without solving an inverse problem.

Efficiency of the method and the system of parameters find another justification in quantitative predictions of spectra measured under different experimental conditions. Here we consider spectra of dispersed fluorescence of anthracene [67] and compare them with the spectra simulated with selective excitation to various single vibronic levels (Fig.2.11). The model provides essentially the same level of accuracy as reported above for predictions of conventional spectra, while dispersed fluorescence, being rather sensitive to variations of excitation conditions, involves quite different sets of vibronic transitions in each case.

Thus, the parametric method has proved to be an accurate and efficient technique to tackle direct problem of computing vibronic spectra. It allows to perform quantitative predictions of structure and properties of molecular excited states and radiative vibronic transition probabilities, including fine spectral effects associated with various substitutions (methyl-, phenyl-, deutero-). These results are found in good agreement with a number of independent spectral experiments and, thereby, confirm possibility of predictive calculations based on the use of small (3-4 atoms) characteristic molecular fragments. Such building blocks (or mini-models) can be calibrated on different levels of approximation depending on the complexity of system under consideration, but the number of relevant, physically reasonable parameters is expected to remain small. This latter factor is believed to be a key to development of specialized databases of molecular fragments that store information accumulated through solution of inverse spectral problems.

## 3. Modeling time-resolved vibronic spectra

It appears quite natural to develop the parametric approach towards modeling time-dependent spectra, which apparently bring more information as compared to conventional vibronic spectra. At first sight, it may seem, though, that all one need to compute dynamical spectrum are the probabilities of vibronic transitions $w_{ij}$ that can be estimated by the parametric method described in previous sections. However, in practice, a number of questions and problems arise.

First, to what extent the parameters of molecular model obtained (calibrated) with the use of conventional (stationary) spectra will be good for modeling time-resolved spectra?

Second, the effect of environment (molecular interactions) may substantially alter the picture seen in spectral experiment with time resolution with the problem being to distinguish between contributions from non-radiative and optical transitions.

Third, absorption, as well as radiative and non-radiative relaxation processes can induce isomerization, which is an issue of particular interest in photochemistry. In part, this problem is traced back to calculating probabilities of optical and non-radiative inter-isomer transitions.

Fourth, in regard to purely computational aspects, besides the fact that an extensive number of transition probabilities $w_{ij}$ have to be calculated, even in case of mid-size molecules ($\approx$ 20–30 atoms) a high dimensional ($N > 10^3$) system of differential equations is to be solved for time-dependent populations $n_i(t)$. Although general methods are well-known, their applicability and performance in large-scale computer simulations are questionable, especially for iterative solution of inverse problems.

Studies in these directions are underway (for recent results see: nonradiative transitions [70], optical inter-isomer transition probabilities [16, 71], generalized inverse problems in vibronic spectroscopy [72]). Here we shall focus primarily on the first part of the problem – direct calculation of time-resolved vibronic spectra for a model of isolated molecule and transitions within single isomer form.

A dynamical time-resolved fluorescence spectrum can be represented as a three dimensional (3D) surface showing dependence of intensity on both frequency and time:

$$I_{ij}(\nu_{ij},t) = h\nu_{ij} \cdot w_{ij} \cdot n_i(t), \qquad (3.1)$$

where $\nu_{ij}$ and $w_{ij}$ are the frequency and probability of transition from the *i*-th to the *j*-th state, $n_i(t)$ is the time-dependent population of the *i*-th state.

Hence, to construct such surface for a given molecular model it is necessary to (i) find transition probabilities $w_{ij}$ for all pairs of $N$ molecular energy levels and (ii) solve the system of kinetic (rate) equations for populations $n_i(t)$. Methods and software for calculation of vibrational and vibronic radiative transition probabilities are well developed [13, 16, 19, 20, 73-75] while the second task suggests that an efficient procedure for solution of high dimensional system of differential equations should be employed since preliminary estimates revealed low performance of standard techniques in real-time computer experiments and for realistic molecular models.

For the set of $N$ excited states sorted according to energy kinetic equations read:

$$\frac{dn_i}{dt} = -w_i n_i + \sum_{j=i+1}^{N} w_{ji} n_j, \qquad i = 1,2,\ldots N, \qquad (3.2)$$

where $w_i = \sum_{j=1}^{i-1} w_{ij} + w_{i0}$ is the total probability (decay rate or inverse lifetime) due to transitions from the *i*-th state to all lower-lying states including the ground state ($w_{i0}$). Initial conditions $\{n_i(0), i = 1,2,...,N\}$ may vary with excitation (e.g. resonance, broadband etc.). Using matrix notation, for the vector of populations $\mathbf{n}(t)$, the system (3.2) can be rewritten as follows:

$$\frac{d}{dt}\mathbf{n}(t) = P\mathbf{n}(t),$$

where $P$ is the superdiagonal matrix of probabilities $p_{ij} = \begin{cases} -w_i, & i = j \\ w_{ji}, & i < j \\ 0, & i > j \end{cases}.$

As analysis of applicable computational techniques has shown [76], for simulation of spectra approximate numerical methods are preferred. Here we proceed with simple first order Euler scheme [77] and obtain the map:

$$\mathbf{n}(t_{k+1}) = \mathbf{n}(t_k) + \Delta t \cdot P \cdot \mathbf{n}(t_k), \qquad t_{k+1} = t_k + \Delta t.$$

Indeed, the speed of calculations according to this procedure scales as $N^2$ and there is no need to store any additional matrices except $P$. Besides, the algorithm can be optimized to gain in speed by a factor of about 100 by using explicitly the triangular form of matrix $P$ and eliminating equations with trivial solutions once initial conditions are specified. But it is more important that numerical solution can be easily obtained for any matrix of probabilities (including one with degenerate eigenvalues that may appear in applications quite often). Moreover, the method becomes equally applicable to systems with time-dependent probabilities, as is the case when quantum beats associated with isomerization are to be taken into account.

Some conclusions immediately follow from the analysis of kinetic equations [76].

The time dependence of population of the *i*-th energy level and, hence, intensities of transitions originated from this level, will show multi-exponential behavior determined by the total probability $w_i$ and those of higher levels $w_k$ ($k>i$). Note that the system of molecular energy levels can be split into groups of vibrational sublevels that belong to different electronic states. Consider the ground state (0) and the first excited state (1) and their vibrational sublevels denoted by the symbols *f* and *i* (these indices refer to sets of quantum numbers). Then the total probabilities of optical vibronic transitions from sublevels of the excited state to sublevels of the ground state (here the probabilities of nonradiative vibrational transitions within excited state manifold are assumed to be zero) are given by:

$$w_{1,i} = \sum_f w_{1,i \to 0,f} = \sum_f \frac{64\pi^4 \nu^3_{1,i \to 0,f}}{3h\varepsilon_0 c^3} \cdot \mu^2_{1,i \to 0,f},$$

where $\omega_{1,i \to 0,f}$ is the frequency and $\mu_{1,i \to 0,f}$ is the dipole moment of vibronic transition.

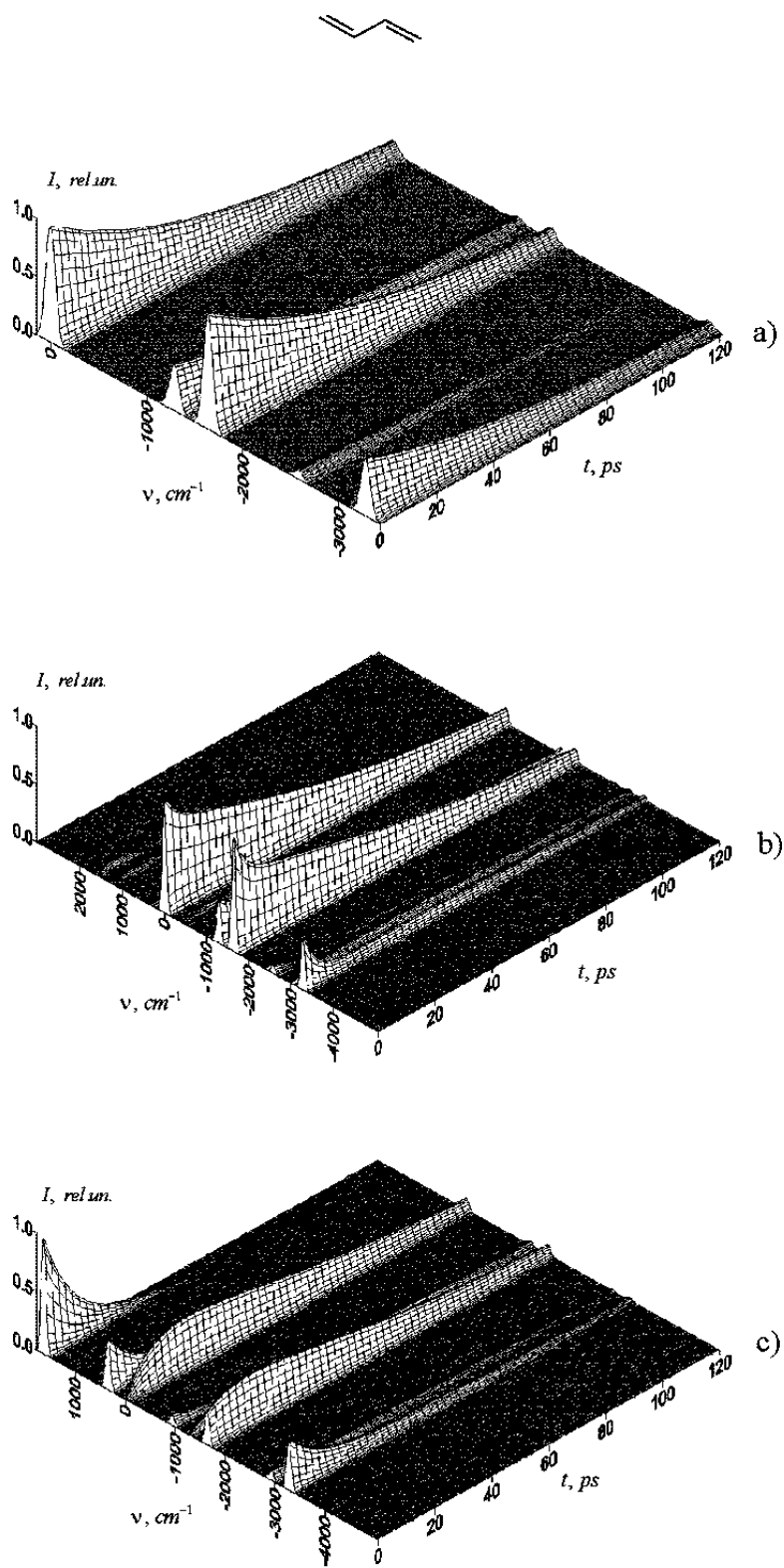

Fig.3.1. Spectra of butadiene calculated with zero (a) and nonzero (b, c) probabilities of nonradiative vibrational relaxation (resonance excitation of vibrational mode 3000 cm$^{-1}$ (a, b) and 1620 cm$^{-1}$ (c) of $S_1$ electronic state). The probabilities of vibrational relaxation are 60% of electronic one.

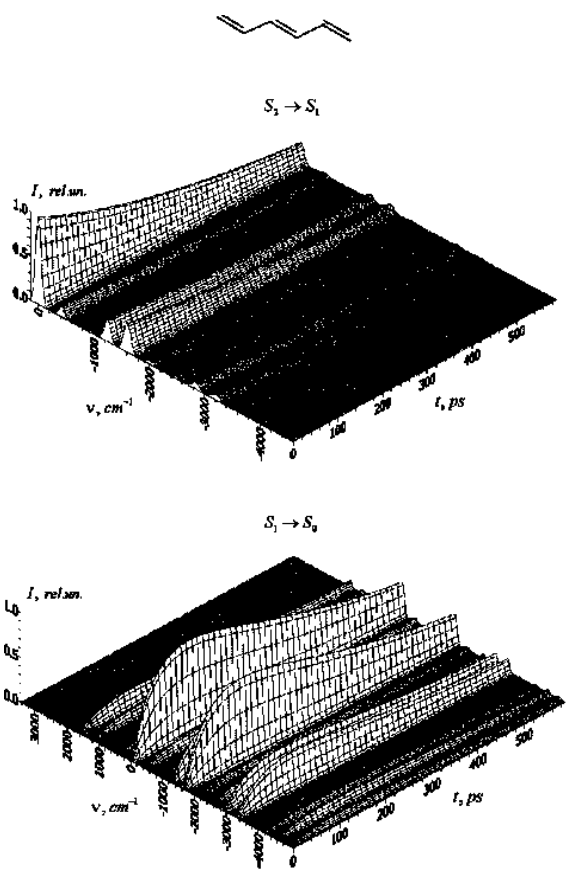

Fig.3.2. Time-resolved fluorescence spectra of hexatriene upon excitation to $S_2$ electronic state.

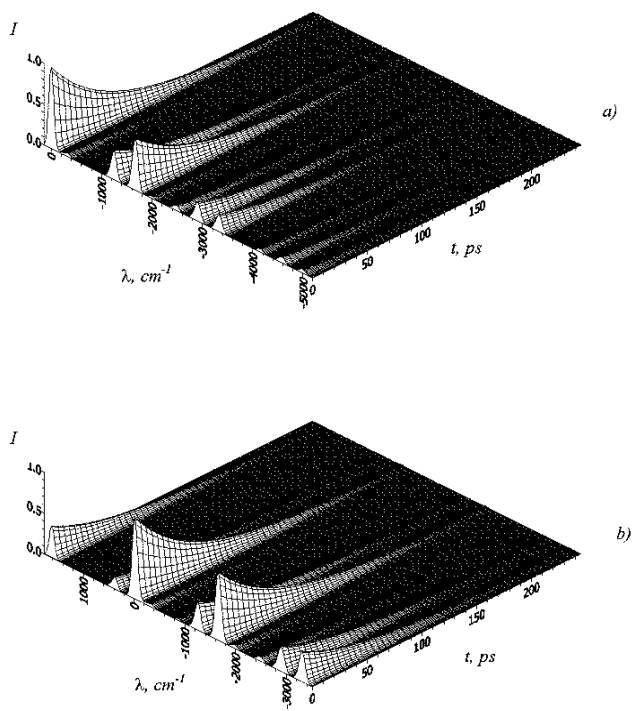

Fig.3.3. Time-resolved fluorescence spectra ($S_1 \to S_0$) of octatetraene upon excitation to $S_1$ (a) and $S_1$+1620 cm$^{-1}$ (b) vibronic states.

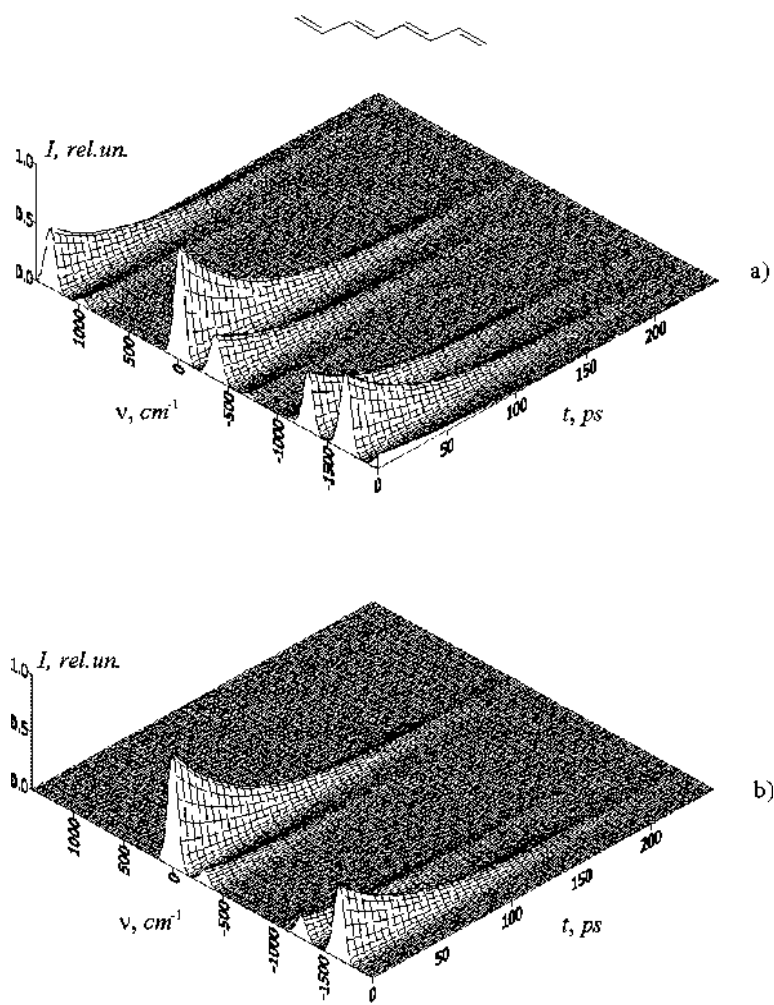

Fig.3.4. Simulated dynamical spectra of octatetraene under resonance excitation of totally symmetric mode $v = 1283$ cm$^{-1}$ (a) and nontotally symmetric overtone mode $v = 1276$ cm$^{-1}$ (b) of the first excited electronic state $S_1$.

In Franck-Condon approximation these probabilities are proportional to the overlap integrals of vibrational wave functions $\langle i|f\rangle$ and for small changes in geometry upon excitation, which are typical for polyatomic molecules, we obtain:

$$w_{1,i} = \frac{4}{3\hbar\varepsilon_0 c^3}\cdot\mu_{10}^2\cdot\sum_f \omega_{1,i\to 0,f}^3 \langle i|f\rangle^2 \approx \frac{4}{3\hbar\varepsilon_0 c^3}\cdot\omega_{10}^3\cdot\mu_{10}^2\cdot\sum_f \langle i|f\rangle^2 = \frac{4}{3\hbar\varepsilon_0 c^3}\cdot\omega_{10}^3\cdot\mu_{10}^2. \quad (3.3)$$

where $\omega_{10}$ and $\mu_{10}$ are the frequency and electronic dipole transition moment.

One can see from equation (3.3) that total probabilities for all vibrational sublevels of an excited electronic state of isolated molecule are the same as determined by values of $\mu_{10}$ and $\omega_{10}$ for electronic transition. It was also verified experimentally (see [78]) with the use of time-dependent intensities of dispersed fluorescence spectra. So, to a first approximation (probabilities of vibrational transitions are negligibly small), all lines within a single electronic spectra, say, $S_1 \to S_0$, will exhibit uniform multi-exponential dynamics, but their intensities will still depend on individual probabilities of vibronic transitions $w_{1,j\to 0,k}$. The implication here is that any differences in time profiles can serve as signatures of vibrational relaxation.

If high vibrational levels (overtones and combination bands) are selectively excited, relaxation may include transitions with almost equal probabilities (particularly those that would have the highest intensity with the change in vibrational quantum number of 1). In this situation the system of kinetic equations does not split into separate blocks but has degenerate (or close to degenerate) eigenvalues and general solution is sought. Partly for this reason, we have built all our algorithms and software making use of numerical integrator rather than attempting to derive an analytical solution in every single case beyond the simplest, easily diagonalizable system. As an additional reward, this strategy provides maximum flexibility for further extensions of the model.

Shown on Fig.3.1-3.3 are some examples of simulated time-resolved spectra for butadiene, hexatriene and octatetraene when various vibronic states (purely electronic, totally symmetric and non-totally symmetric vibrations) are initially excited. Spectral patterns that correspond to different symmetries of vibrational wave function appear qualitatively distinct. Even small deviation in excitation frequency ($\Delta\nu \approx 10\,cm^{-1}$) is detectable (see Fig.3.4) by comparing relative intensities of strong lines ($\nu \leq 0$) and by the absence of signal for frequencies above the electronic origin ($\nu > 0$). This, in particular, gives an estimate for the selectivity of excitation required to produce a good experimental spectra that can be compared with simulations or used for solution of an inverse problem. It is also notable that excitation of totally symmetric vibrations gives rise to more informative spectra and simulations can help in the search for conditions which would be optimal for a specific experiment pursuing quantitative or qualitative analysis (for example, frequency of the most active mode can be estimated).

What might also turn out to be nontrivial is the issue of finding appropriate spectral and temporal ranges for experimental measurements if the set of high-energy molecular levels is excited by a broadband pulse. Under such conditions decay evolves through a number of intermediate vibronic states so that fluorescence spectrum is expected to be quite complicated. For example, as shown on Fig.3.5, the full time-resolved spectrum of diphenylbutadiene excited into $S_5$ electronic state consists of strong band due to transition $S_5 \to S_0$ and substantially (approximately by a factor of 100 and 1000, respectively) less intense components $S_1 \to S_0$ and $S_4 \to S_1$ (all the rest transitions have zero probabilities). Modeling can help determine that, in this case, the properties of intermediate $S_1$ and $S_4$ states and transition probabilities can be retrieved (through solution of inverse spectral problem [79, 80]) from weak and slowly changing spectra of transitions $S_1 \to S_0$ (Fig.3.5d) and $S_4 \to S_1$ (Fig.3.5e) measured within the spectral ranges $29000 \div 36000\,\text{cm}^{-1}$ and $4000 \div 10000\,\text{cm}^{-1}$ correspondingly with required time being approximately $t_{observ} \approx 2500\,\text{ps}$. Note that it could be difficult to make such estimates just left with a full experimental spectrum like shown on Fig.3.15 alone and without simulations, cause due to low sensitivity or insufficient spectral/temporal resolution details of weak $S_1 \to S_0$ and $S_4 \to S_1$ transitions could almost completely be hidden under the strong $S_5 \to S_0$ (Fig.3.15c) band. Thus, model calculations can effectively guide searches for spectral regions that contain "trace amounts" of information on structure and dynamics of intermediate states.

For solution of an inverse problem it is important to have as detailed experimental spectrum as possible which presumably should depend on most parameters (transition probabilities) to be determined at once. Computer experiments confirm strong dependence of dynamical spectrum on excitation conditions and what immediately follows from this is the possibility of controlling 3D spectral signal for a given molecule. For example, if $S_1$ electronic state is excited, $S_1 \to S_0$ spectrum (Fig.3.6a) apparently gives only probabilities of transitions originating from the vibrational sublevels of $S_1$ manifold. Alternatively, once a molecule is excited by two synchronous laser pulses ($S_1$ and $S_5$ are initially populated), this same spectrum now depends on all nonzero probabilities of transitions between states below $S_5$ and ratio of intensities of excitation pulses $I_1$ and $I_5$. At $I_5/I_1 = 100$ this effect is still quite small (Fig.3.6b), but under optimal conditions ($I_5/I_1 = 1000$, Fig.3.6c) multi-exponential profile of $S_1 \to S_0$ dynamical spectrum reflects explicitly the total probabilities (decay rates) of all three excited states $S_1$, $S_4$ and $S_5$ ($S_3$ and $S_2$ do not contribute due to zero probabilities of $S_5 \to S_3$, $S_4 \to S_3$, $S_5 \to S_2$ and $S_4 \to S_2$ electronic transitions). Note that such control over spectral picture comes from the use of additional pulses, while simple excitation to high electronic states ($S_5$, Fig.3.5d) does not help — $S_1 \to S_0$ spectrum is almost entirely determined by the lifetimes of $S_4$ and $S_5$ and does not clearly show decay of $S_1$ (corresponding weak component can be traced only over relatively short times, $t < 80$ $ps$). Here the role of the second pulse is to strengthen this weak band making it possible to extract its probability from experimental spectrum.

It is also essential that direct simulations of dynamical spectra together with experimental data and methods for solution of inverse problems can eventually provide insights into probabilities of nonradiative vibrational processes in excited electronic states. As mentioned above, vibrational relaxation manifests as differences in time dependences of individual vibronic lines. A quantitative approach to this problem and technique that would allow for possibility of obtaining the values of probabilities from spectral data suggests certain experimental setting that makes the effects associated with nonradiative dissipation reliably detectable. Consider the case when probabilities of vibrational transitions $w^{vibr}$ are comparable with those of electronic transitions $w^{el}$. Due to sensitive dependence of time-resolved spectrum on variation of excitation conditions, with increasing energy of excitation pulse spectrum evolves from almost identical to that calculated with $w^{vibr} = 0$ (compare Fig.3.7a and Fig.3.7b) toward complex nonuniform (Fig.3.7d) and multiexponential (Fig.3.7f) structures. Such experimental spectra obtained with broadband excitation to high-energy sublevels carry more information about relaxation processes and can be more effectively used. Indeed, the possibility to extract the values of nonradiative transition probabilities from 3D spectrum comes from the fact that intensities depend on different total probabilities $w_i = w^{el} + \sum_j w_{ij}^{vibr}$. In turn, $w_{ij}^{vibr}$ are defined by the parameters of molecular model (in particular, by the intermolecular potential [70]). So, direct variation of these parameters can give the values of $w_{ij}^{vibr}$ (along with variable parameters themselves) as the calculated spectrum approaches experimental one in the course of iterations. The amount of information contained in the experimental spectrum, thus, clearly correlates with the number of spectral lines that exhibit various behaviors.

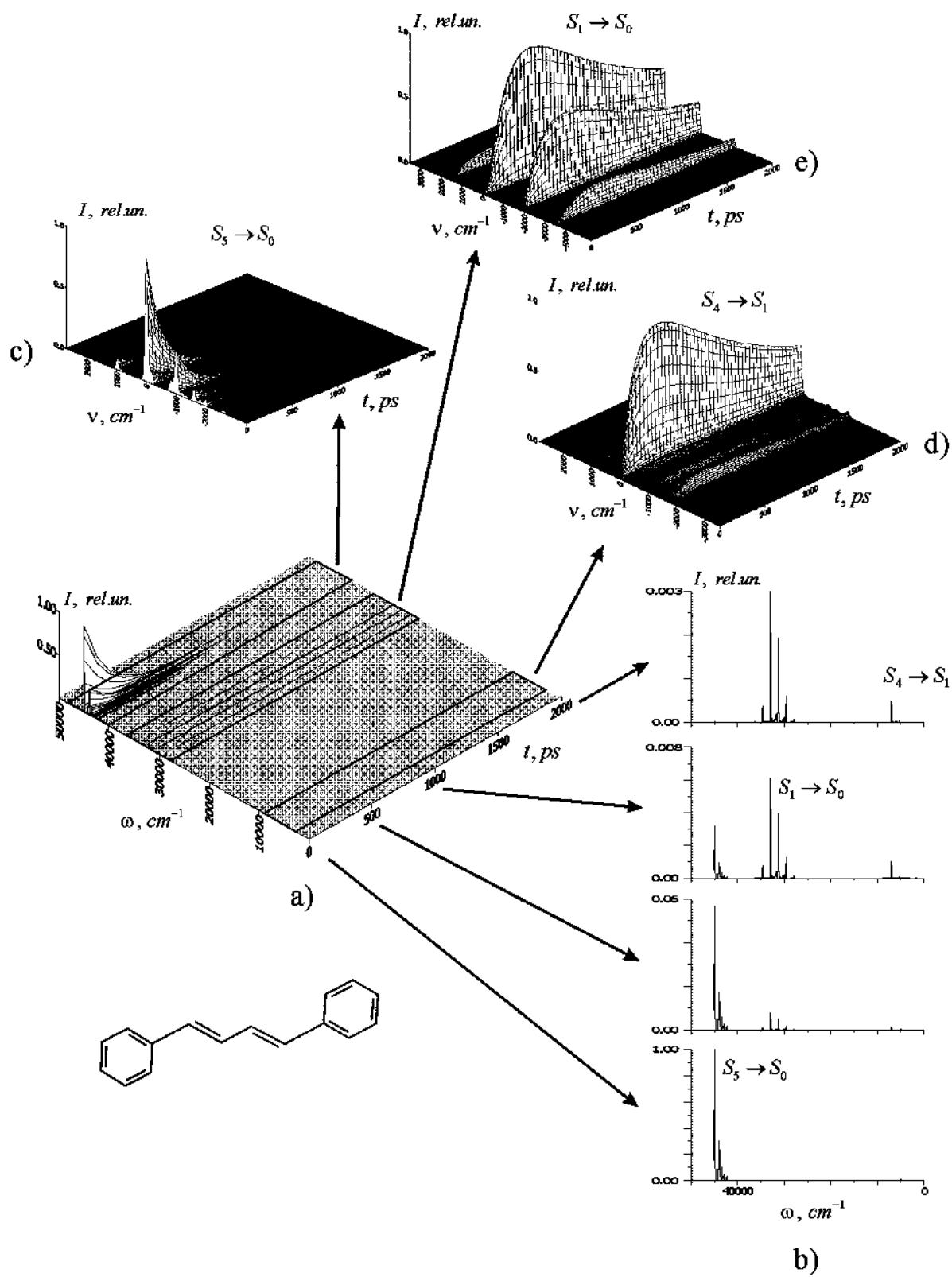

Fig.3.5. Full-scale calculated spectra of diphenylbutadiene under excitation of all vibrational sublevels of $S_5$ electronic manifold (a), its sections at $t = 0, 500, 1000, 2000$ ps (b) and spectral regions that correspond to different electronic transitions (c, d, e).

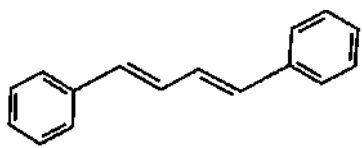

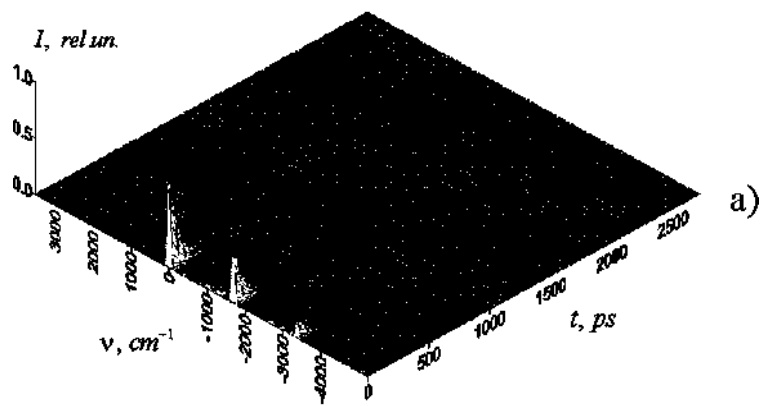

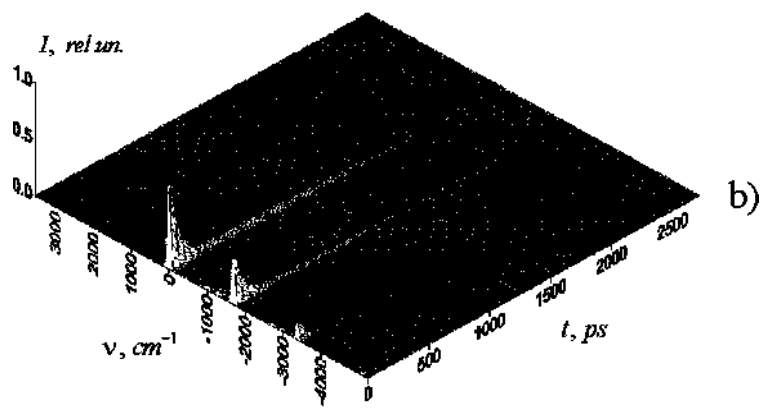

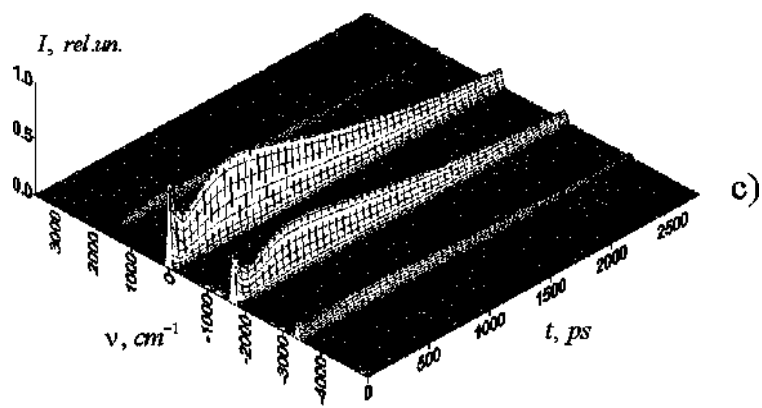

Fig.3.6. Spectra of $S_1 \rightarrow S_0$ transition in diphenylbutadiene upon excitation to $S_1$ state (a), $S_1$ and $S_5$ states simultaneously with relative intensities $I_5/I_1 = 100$ (b) and $I_5/I_1 = 1000$ (c).

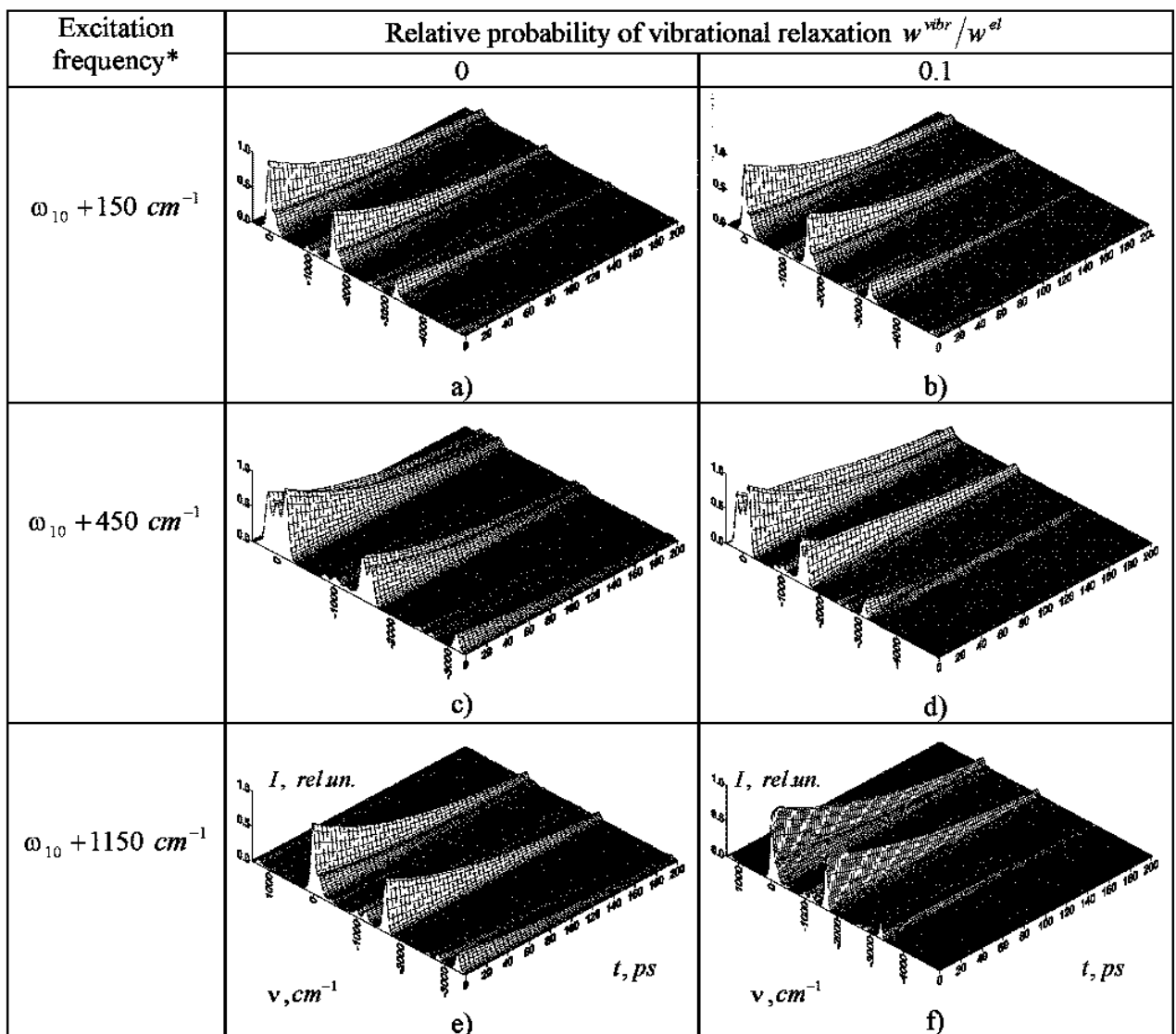

Fig.3.7. Dynamical spectra of stilbene upon wide-band excitation of vibrational sublevels of $S_1$ electronic state calculated with zero probability of vibrational relaxation (a, c, e); (b, d, f).— the case when probability of vibrational relaxation is comparable ($w^{vibr}/w^{el} = 0.1$) with electronic one.

Thereby, simulations of dynamical vibronic spectra being a window into developing quantitative models of nonradiative processes, can help extend methods of quantitative and qualitative standardless analysis based on time-resolved spectroscopy [81, 82] to the case of dense media (solid gases, liquids) where the effect of collisional nonradiative intermolecular relaxation is important, since molecular interactions irreversibly transfer some part of excitation energy into external (translational) degrees of freedom that results in increase of ensemble temperature. Quantitatively, the probabilities of nonradiative transitions $w_{ij}^{NR}$ appear in expressions for time-dependent populations and fluorescence intensities $I_{ij}(\omega_{ij},t) \sim w_{ij}^{NR} \sum_k A_{ik} e^{-w_k t}$, where total probabilities now consist of radiative and nonradiative components $w_k = w_k^R + w_k^{NR}$, whereas coefficients $A_{ik}$ depend on individual transitions probabilities $w_{lm} = w_{lm}^R + w_{lm}^{NR}$. Unfortunately, methods of inverse spectral problems do not give an unambiguous solution for unknown $w_{ij}^{NR}$. Part of the reason is that these quantities, even if determined for a given molecule, do not satisfy the transferability property, and so cannot be used for modeling spectra of different molecules of the same family or in different experimental conditions. Moreover, having measured relative intensities $I_{ij}(\omega_{ij},t)/I_{ij}(\omega_{ij},t') = \sum_k a_{ik} e^{-w_k t}$, where $a_{ik} = A_{ik}/B_i$, $B_i = \sum_k A_{ik} \exp(-w_k t')$, one can find parameters $A_{ik}$ only accurate to constants $B_i$ and, hence, quantities $w_{ij}^{NR}$ as well as $w_{ij}^R$ are accurate to $B_0$, which corresponds to some reference line in the spectrum, because in a spectral experiment the fraction of excitation energy transferred to thermal modes due to collisions is left unknown. This uncertainty can be resolved either by using spectra with absolute intensities, although it is difficult in practice, or via an additional measurement of change in temperature or other thermodynamical parameters of sample (for a review on photothermal spectroscopy see [83]). We shall also discuss nonradiative transitions in the next section.

Quite often, features of conventional spectra (without time-resolution) of a number of molecules (for example, polyenes, phenyl- and diphenylpolyenes), their stereoisomers or substituted forms (*cis*, *trans*-isomers, deutero-, methyl-, phenyl-substituted) are almost identical [16, 48-51, 63-66, 84-86]. In this situation differences in intensities of major vibronic lines (~10%) may fall, in fact, within the error bar of calculations that makes interpretation of such spectra, much less identification of compounds (especially in a mixture), nearly impossible. This obstacle significantly restricts the use of some spectral methods in analytical chemistry and photochemistry (in particular, if photoisomerization is involved [33]).

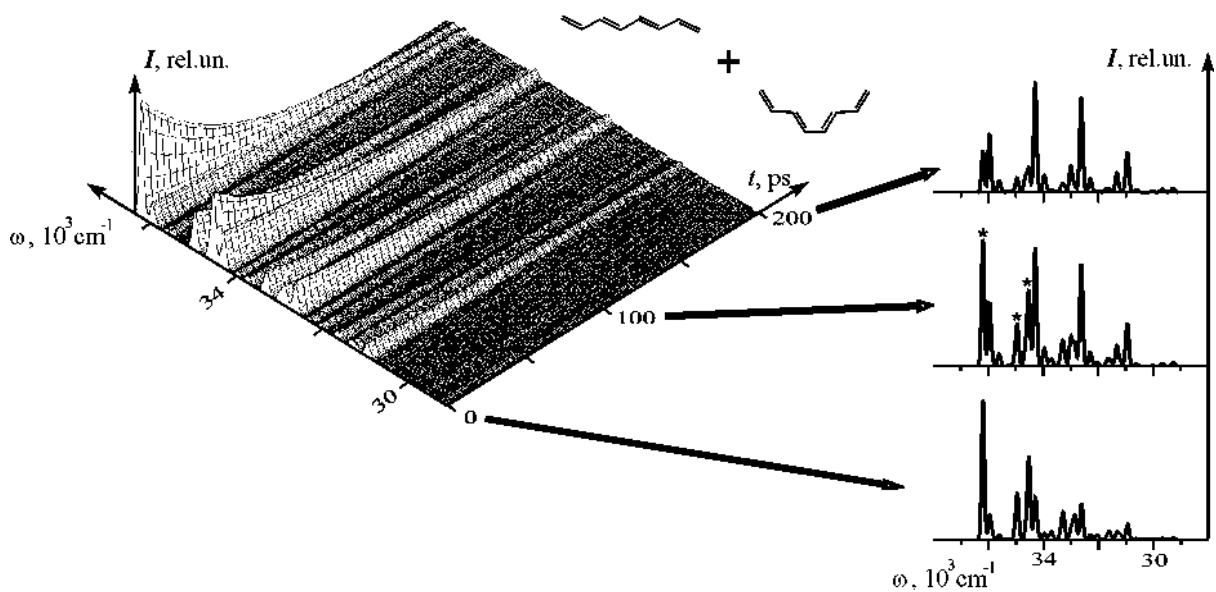

Fig.3.8. Calculated time-resolved fluorescence spectrum of the mixture of rotational isomers of octatetraene.

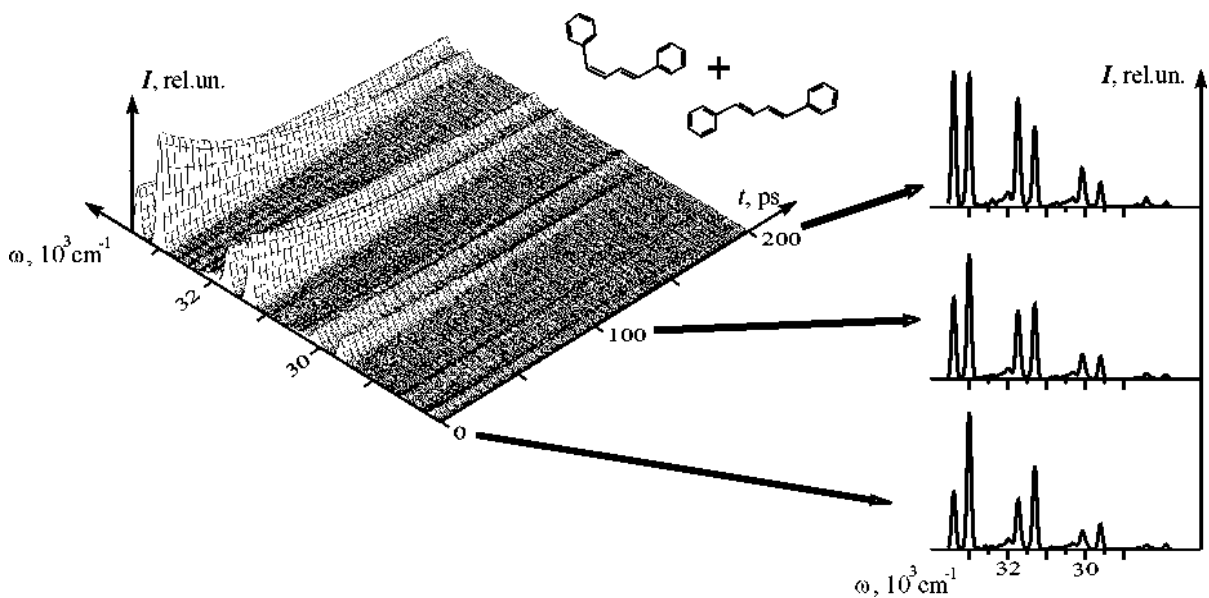

Fig.3.9. Calculated time-resolved fluorescence spectrum of the mixture of rotational isomers of phenyl-substituted butadiene.

The approaches from time-resolved spectroscopy provide a technique to tackle these problems in a more efficient way. Shown on Fig.3.8 is 3D spectrum of mixture of *cis*- and *trans*-octatetraene and its sections at different times. It is clearly seen that evolution of spectrum suggests that contributions from different components of mixture can be distinguished (assigned) based on their unique time-dependences (those strong bands that correspond to *trans*-isomer are marked with asterisks). It was verified in [81] that such characteristic differences in time profiles allow to perform both qualitative and quantitative analyses in a wide range of relative concentrations (in this case — from $C_{cis}/C_{trans} = 0.45$ up to $C_{cis}/C_{trans} = 45$).

Note also that while energies of purely electronic transitions of isomers differ only slightly, by one or two vibrational quanta (~2000 $cm^{-1}$ for *cis*- and *trans*-octatetraene), simultaneous excitation of mixture components by even spectrally narrow pulse produces quantitatively distinct distributions of initially populated vibronic states at each of the isomers. For instance, the spectrum on Fig.3.8 represents dynamics occurring when purely electronic state of *trans*-octatetraene and 10 sublevels (totally symmetric vibrational modes with energies $0 < \omega < 2000\ cm^{-1}$) of *cis*-octatetraene are initially excited. This gives the presence of signal in the range above the electronic origin for *cis*-octatetraene as another signature of this component in the mixture.

The use of dynamical spectra can be equally efficient also in the case when the problem is to identify molecules with different locations of substituents and it is difficult to do that using spectra without time resolution. For *cis*- and *trans*-diphenylbutadiene computed differences are typically less than 10% for frequencies, intensities in vibronic spectrum and energies of electronic transitions (33400 and 33000 $cm^{-1}$). This can be traced to almost identical structure of polyene chains in such compounds. In addition, as calculations have shown, changes in electron density upon transition to the first excited state are well localized mainly within the chains rather than phenyl rings. For this reason, bond lengths and force constants in "polyene fragments" show 2-3 times greater changes (as compared to phenyl rings) specifically responsible for the structure of vibronic spectrum and its independence on locations of substituent groups. At the same time, these species can be identified by characteristic decay profiles in time-resolved spectrum (see sections on Fig.3.9), even though bands overlap each other. It was also confirmed by computer experiments that efficiency of such recognition procedure increases as molecules are excited to higher electronic states.

The researches summarized herein have been aimed to show some new potentials and efficiency of approaches from time-resolved spectroscopy as estimated by means of simulations. In this still developing field computer experiments and molecular modeling [76, 81, 87] can be very useful, able to guide experiments with an appropriate choice of optimal conditions and settings that would otherwise be hard or extremely resource-consuming to find in a purely empirical manner.

## 4. Simulations of dynamical spectra and inter-isomer transitions

So far, the model of isolated molecule and transitions in the single isomer form were assumed. To further extend applications of computational methods described in previous sections, we address the problem of computing dynamical spectra when transitions between different isomers are taken into account. Together with methods developed for calculation of photoinduced inter-isomer transition probabilities [18-20, 48-50, 73, 74, 82], direct simulations of dynamics provide an approach to mechanisms and rates of isomerization in polyatomic molecules.

We will make use of the theoretical model of isomerization proposed earlier [88]. External perturbation imposed on resonance levels of isomers ( $E = E_1 = E_2$ ) with nonzero coupling (nondiagonal element of energy matrix) gives rise to quantum beats, i.e. periodic oscillations in concentrations of isomers. This superposition state is described by the wavefunction $\psi(t) = \psi_1 \cos\omega t + i\psi_2 \sin\omega t$ or probability density:

$$|\psi(t)|^2 \sim (\psi_1^2 \cos^2 \omega t + \psi_2^2 \sin^2 \omega t), \qquad (4.1)$$

where oscillations of probability of finding the system in pure isomer states $\psi_1$ and $\psi_2$ follow $\cos^2 \omega t$ and $\sin^2 \omega t$ correspondingly. The frequency of quantum beats is defined by nondiagonal matrix element (doublet splitting):

$$h_{12} = \frac{1}{2}(E_1 + E_2)S_e^{(1,2)}S_v^{(1,2)},$$

where $S_e^{(1,2)}$ and $S_v^{(1,2)}$ are the overlap integrals of electronic and vibrational wavefunctions of isomers, respectively (Franck-Condon approximation), $E_1$ and $E_2$ are vibronic energies in the zeroth approximation. Methods for calculation of matrix elements and $\omega$ have been developed in [18-20, 48-50, 73, 74, 88].

The wavefunction $\psi$ appears time dependent, although in the following the energy levels are assumed stationary:

$$H = \int \psi^* \hat{H} \psi dv = \left(\int \psi_1 \hat{H} \psi_1 dv\right)\cos^2 \omega t + \left(\int \psi_2 \hat{H} \psi_2 dv\right)\sin^2 \omega t = E_1 \cos^2 \omega t + E_2 \sin^2 \omega t = E,$$

where time-independent Hamiltonian $\hat{H}$ acting on wavefunctions of isomers 1 and 2 yields the energies of resonant states, $\hat{H}\psi_1 = E_1\psi_1$ and $\hat{H}\psi_2 = E_2\psi_2$. As a result, the model represents a set of stationary levels and optical transitions to and from common resonance level with probabilities defined by dipole transition moments $\left(\langle\mu\rangle_j^i\right)^2$ multiplied by $\cos^2 \omega t$ or $\sin^2 \omega t$.

If, for example, an excited state of isomer 1 is initially populated, then decay evolves through two channels (isomers 1 and 2) with different rates. Some partion of excited molecules will, thereby, eventually undergo transition to the second isomer form. The system of kinetic equations will have essentially the same form, (3.2), but some of the probabilities $w_{ij}$ will be time-dependent, $w_{ij} \cos^2 \omega t$ or $w_{ij} \sin^2 \omega t$, in every case where one of the combining states, $i$ or $j$, belongs to isomer-isomer resonance described by (4.1). The numerical technique and software discussed in the previous section are equally applicable to solution of time-dependent kinetic equations and computing dynamical spectra of coupled isomers. Here we outline some examples of numerical experiments.

Resonant mixing of stationary levels of isomers (Fig.4.1) and quantum beats manifest as oscillatory behavior of intensities in time-resolved spectra (Fig.4.2). For this set of simulations the initial conditions were set so that only levels of isomer 1 are populated at $t = 0$, i.e. $n_1(0) = n_2(0) = 1$, $n_3(0) = n_4(0) = 0$. Transition probabilities were chosen to be $w_{12} = 0.9 w_{13}$, $w_{23} = 0.7 w_{13}$ and $w_{24} = w_{23}$. Intensities of $v_{12}$ and $v_{23}$ lines are directly related to resonance and demonstrate distinct oscillations. Note also how quantum beats, originated from resonance level 2, indirectly affect the time-dependent intensity of $v_{13}$ line. The amplitude of these oscillations is defined by the proportion between $w_{13}$ and $w_{12}$. Transition $1 \to 2$ contributes also to $v_{23}$ band, whose intensity initially increases $(w_{12} > w_{23})$, i.e. transition to isomer 2 occurs. With this transformation the growth and consecutive $\sin^2 \omega t$ like dynamics of $2 \to 4$ spectrum are associated (Fig.4.2b). The quantum yield of isomerization is related to asymptotic ratio $n_4(t)/n_1(0)$ at $t \to \infty$, which is about 20% in this particular case.

As a second example consider model calculations of isomer transformation from pentadiene-1,3 to pentadiene-1,4 under various initial conditions. These simulations were aimed to investigate the interplay between quantum beats and exponential decay of fluorescence intensities for realistic molecular models. We have determined that possible resonance may exist between 1661 cm$^{-1}$ (pentadiene-1,3) and 1293 cm$^{-1}$ (pentadiene-1,4) sublevels of the first excited states, since these vibrational modes most closely match specific structural transformations of these molecules in the course of isomerization. The plots presented here show dynamical spectra upon broadband excitation (Fig.4.3-4.5) and selective excitation to 1661 cm$^{-1}$ sublevel (Fig.4.6) of pentadiene-1,3 for $\omega \gg w$, $\omega \approx w$ and $\omega \ll w$, where $w$ stands for typical probability of single isomer radiative transition (all intensities were normalized by their maximum values).

These results indicate that while at $\omega \gg w$ (Fig.4.3) and $\omega \approx w$ (Fig.4.4) spectra of two isomers have nearly equal intensities, in the opposite case (slow quantum beats, $\omega \ll w$, Fig.4.5) signal from isomer 2 comes out 4 orders of magnitude lower than that of isomer 1. This means that such weak spectrum will be almost completely masked by overlapping strong bands of isomer 1. It was estimated that spectrum of isomer 2 can be reliably detected (with intensity at least 10% relative to isomer 1) if $\omega > 0.7w$. Low intensity of fluorescence from isomer 2 signifies low relative number (concentration) of molecules that underwent isomerization (this quantity approaches zero when $\omega \ll w$, but it is still 1.5% at $\omega \gg w$ and 2.0% at $\omega \approx w$). Consequently, slow quantum beats have no effect on dynamics of spectrum of isomer 1 (Fig.4.5a), as compared to the case of isolated molecules. Note, however, that fluorescence signal from isomer 2 can still be observed if it falls within slightly different spectral range and does not overlap with dominant spectrum of the first isomer.

Spectro-temporal patterns observed when $\omega \gg w$ (Fig.4.3) and $\omega \approx w$ (Fig.4.4) are similar and differ only quantitatively, mainly in frequency of quantum beats which appears explicitly in all time dependences. Spectrum of isomer 2 is fully modulated (its intensities approach zero at some points), while spectrum of the first isomer looks evidently more complex. This is due to not all the spectral lines exhibit oscillations; there are some that have nothing to do with resonance but contribute to superposition. It becomes clear when resonance level is exclusively selected by excitation (Fig.4.6) and all non-oscillatory contributions vanish. Dynamics of populations of the first isomer, its spectral response and quantum yield of isomerization are rather sensitive to variation of excitation conditions. But, as long as isomers are assumed coupled through the single resonance level, normalized spectrum of the second isomer does not change upon these variations (see Fig.4.4b and Fig.4.6b).

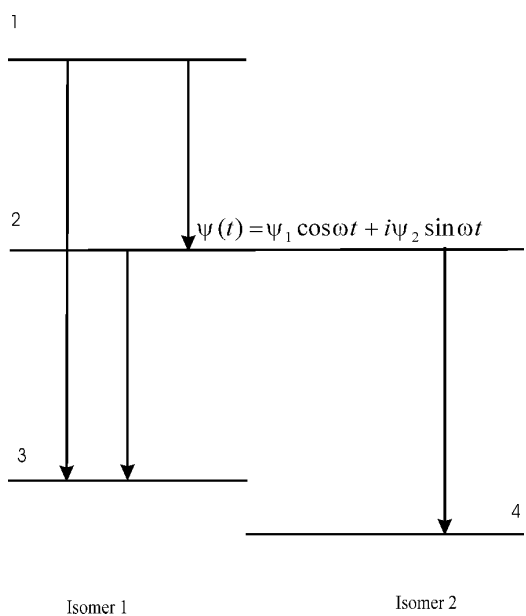

Fig.4.1. Energy levels and transitions due to resonance interaction between molecular isomers; emerging "isomer-isomer" state.

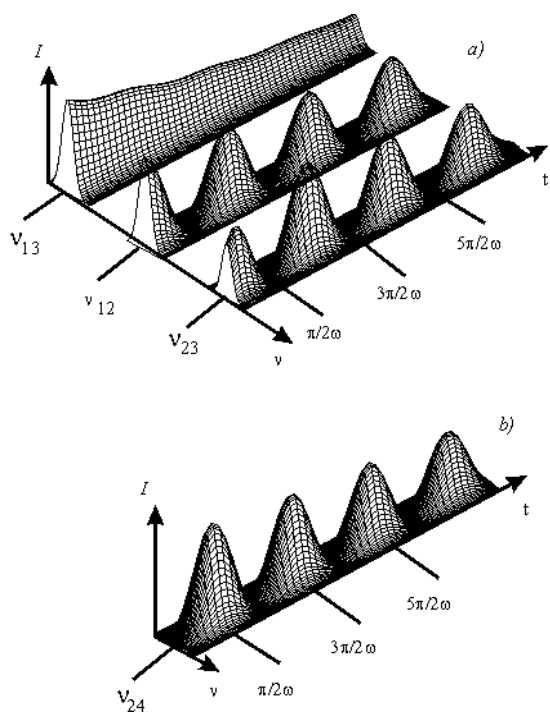

Fig.4.2. Fluorescence spectra of isomer 1 (a) and isomer 2 (b) for the model system (Fig.4.1).

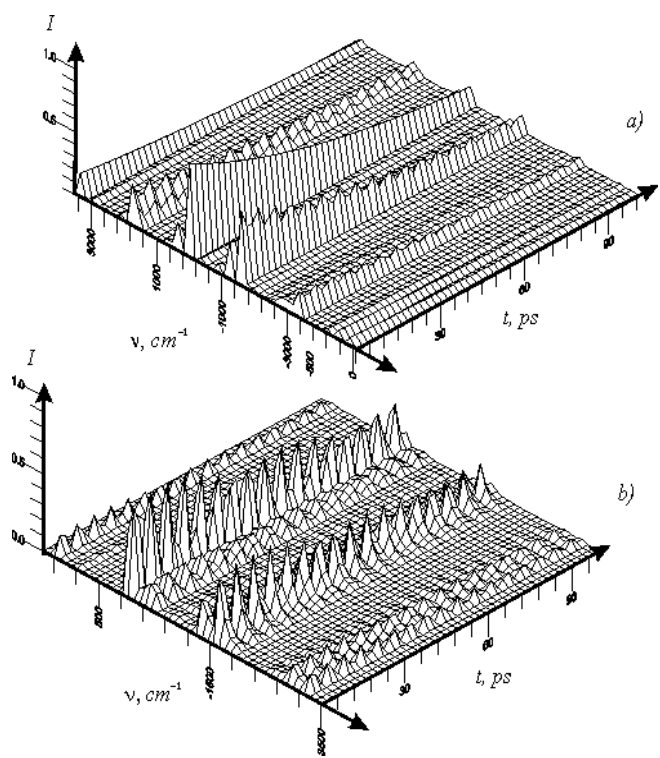

Fig.4.3. Fluorescence spectra ($S_1 \rightarrow S_0$) of pentadiene-1,3 (a) and pentadiene-1,4 (b) upon wide-band excitation of pentadiene-1,3 with inter-isomer transition taken into account and $\omega \gg w$.

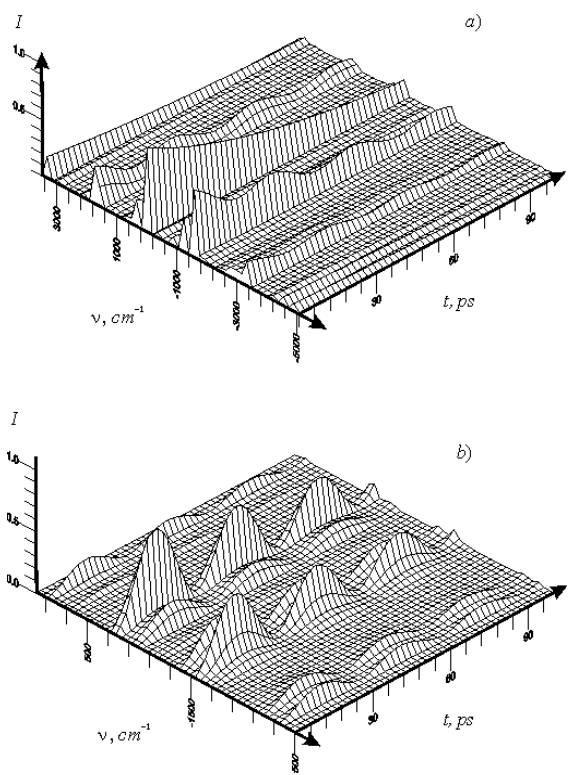

Fig.4.4. Same as Fig.4.3 but $\omega \approx w$.

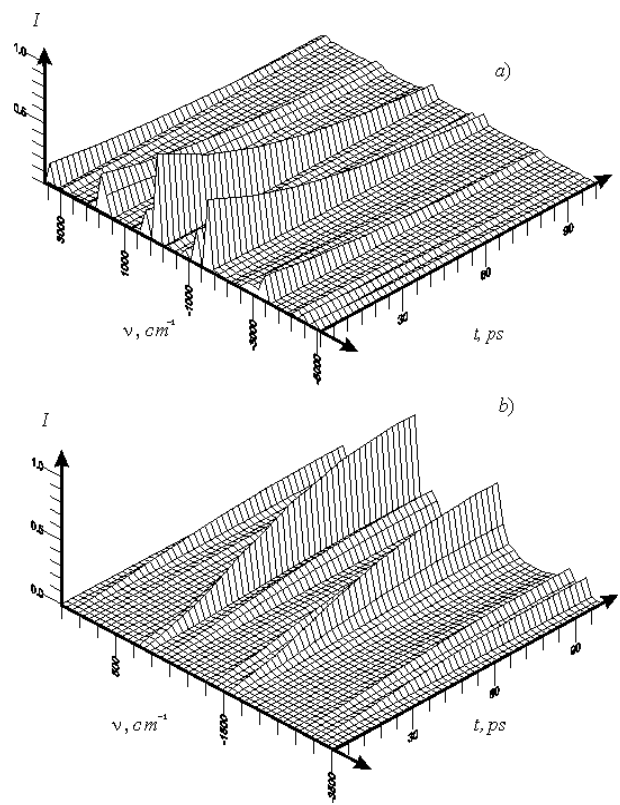

Fig.4.5. Same as Fig.4.3 but $\omega \ll w$.

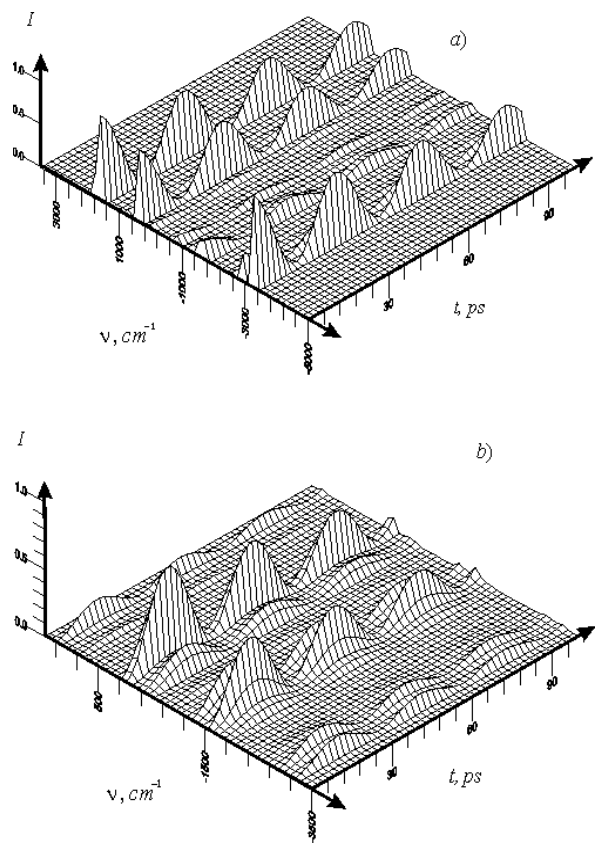

Fig.4.6. Same as Fig.4.4 but upon resonance excitation of pentadiene-1,3.

# 5. Nonradiative transitions in gases under optical excitation

Some of the spectral effects can be attributed to transitions without emitting electromagnetic radiation (nonradiative or radiationless transitions). Among them are: shape of the emission spectrum and fluorescence quantum yield are independent of the excitation wavelength (Vavilov's law [89] for condensed phases or dense gases); fluorescence quantum yield is less than one; for the vast majority of organic molecules fluorescence originates from the lowest excited state of a given multiplicity (Kasha's rule [90]). All these effects are related to ensembles of molecules. However, the common approaches to nonradiative transitions make use of the model of isolated molecule interacting with "bath modes". Such models do not explicitly account for real properties of environment and individual molecules, nor do they provide a description for microscopic details of energy transfer.

In dilute gases transitions between energy levels are necessarily accompanied by emission or absorption of electromagnetic waves (radiative optical transitions). Purely nonradiative transitions are possible only if there is some mechanism for transfer of excess energy to or from the external (with respect to isolated molecule) degrees of freedom. So, we arrive at the model of molecular collisions with relative positions of interacting molecules as extra degrees of freedom. Physically, nonradiative transitions represent the second (alternative to emission) channel for deactivation of the energy of initial optical excitation, which leads to an increase in ensemble temperature. Therefore, such dissipation manifests as a degradation of the total energy (integral intensity) re-emitted by molecules, rather than as a change in emission line shape. Here we pursue the goal of developing theoretical models of nonradiative transitions that would explain the observed effects and provide reasonable (at least semi-quantitative) description for them.

For rare and moderately dense gases, we can rely on the hypothesis of "molecular chaos", i.e. the absence of correlations between colliding molecules, and consider only pair-wise molecular interactions. This model is quite applicable for rare gases, which case we will stick to in the following, when the mean free path $\lambda$ is much greater than effective size of molecules $L$ $(\lambda \gg L)$. For polyatomic gases at normal conditions this requirement holds $(\lambda \sim 10^{-5}\,\text{cm},\ L \sim 10\,\text{Å},\ \lambda \sim 10^2 L)$. Without the loss of generality, we will be content with the model of gas being an ensemble of $N$ identical pairs of (different) molecules A–B. In general, while further considering multicomponent mixtures, one can separate the ensemble into partial subsystems with pair-wise interactions, e.g. for two-component mixtures that would give A–A, B–B and A–B, each of which can be treated by the method described below with summation over all subsystems.

To zero approximation, when internal and translational degrees of freedom are decoupled, the Hamiltonian of pair has additive form:

$$\hat{H}_0 = \hat{H}^A + \hat{H}^B + \hat{H}^T, \tag{5.1}$$

and hence eigenfunctions read:

$$\Psi = \Psi^A \Psi^B \Psi^T, \tag{5.2}$$

where $\hat{H}^A$, $\hat{H}^B$ and $\Psi^A$, $\Psi^B$ are the Hamiltonians and eigenfunctions of internal vibronic degrees of freedom of isolated molecules A and B ($\hat{H}^A \Psi^A = E^A \Psi^A$, $\hat{H}^B \Psi^B = E^B \Psi^B$), $\hat{H}^T$ and $\Psi^T$ represent the operator and wavefunction of relative motion of molecules ($\hat{H}^T \Psi^T = E^T \Psi^T$). Here we speak of translations only and neglect relative rotations with the reason being that, due to rather weak coupling, the probability of collision energy transfer to rotational degrees of freedom appears much lower than to translational ones. Energy re-distribution between translational and rotational modes in thermal equilibrium can then be estimated using the theorem of equipartition of energy among degrees of freedom. Thus, we are left with a single coordinate, namely with the distance $R$ between centers of mass of molecules A and B, describing their relative motion.

Since we are dealing with collisions, translations should be considered in half-bounded space of variable $R$ or in a potential well that acts as a wall form the one side ($R \to 0$, $V \to \infty$) but does not restrict the motion from the other side ($R \to \infty$, $V \to 0$). In addition, molecules at close distances experience van der Waals interactions and the potential has a minimum at some distance $R = R_0$ where the probability of finding the system is an order of magnitude higher than in the region of "force-free approach" ($R \gg R_0$). There exist a number of model potentials that have such properties allowing to describe both relatively long-lived dimer states and isolated molecules of a pair (for example, Lennard-Jones and Buckingham potentials, etc.). To make mathematics simpler we choose Morse wavefunctions with approximate representation being [91]

$$\Psi^T = N_{norm} f(R - R_0) Cos[\omega(R - R_0)], \tag{5.3}$$

with

$$f(R - R_0) = \sum_{l=1}^{n} C_l \exp\left[-\frac{(R - R_0 - l\sigma \ln 2)^2}{\sigma^2}\right]. \tag{5.4}$$

The number of terms in (5.4) $n$ may vary depending on the nature of a problem; the parameter $\sigma$ should be set so that $f(R - R_0)$ approaches unity everywhere except boundaries of the domain of interest, while coefficients $C_l$ come from variational solution of Schrödinger equation with Morse potential. The phase of oscillatory factor in (5.3) is chosen to be zero to maximize the probability to find the system in the centre of the well $(R = R_0)$. The parameter $\omega$ is directly related to the temperature of ensemble: $\omega = \hbar^{-1}\sqrt{kTM}$, where M is the reduced mass of molecules A and B, $k$ is the Boltzmann's constant. $N_{norm}$ is just the normalization factor.

The main difficulty with using functions like (5.3), as well as any other functions that model molecular interactions, is that they cannot be normalized and corresponding spectra is continuous, which is characteristic to motion in half-bounded space. This difficulty can be overcome if we consider that the potential and function (5.3) simulate not just the behavior of two "isolated" molecules A and B, but rather interactions and relative motions of any pair of molecules in the media (gas) averaged over entire ensemble. In this sense, the distances between A and B do not exceed (in average) the mean free path $\lambda$ and, therefore, the domain for the model function (5.3) is defined by $0 < R \leq \lambda$, which is equivalent to modification of the potential of molecular interactions, namely: within $0 < R < \lambda$ it is the Morse potential, but at $R > \lambda$ it goes to infinity. The discontinuity that appears at $R = \lambda$ can be smoothed out in the neighborhood of this point.

We define thereby the normalizable function $\Psi^T$ that corresponds to $\hat{H}^T$ and describes relative translation of a pair of molecules. The appropriate choice of parameters in (5.3) can be obtained by variational method for the Schrödinger equation with a given Morse potential so that energies are determined with ~0.1% error bar [91]. Note, however, that, strictly speaking, the wavefunction (5.3) is not an eigenfunction of any operator. But this fact is by no means a barrier for model approaches because it is not so important whether approximate potential or physically meaningful wavefunction underlies the model (see for example currently developing methods of so-called "intuitive" quantum mechanics [92]).

It is also assumed that intensity of incident radiation is low so that all the interactions are linear. Then for a given energy level there are only "optical" and "thermal" channels of excitation or deactivation and these processes cannot take place simultaneously. In molecular systems transitions between different states are related to matrix elements of corresponding operators. Optical (radiative) transitions are associated with the dipole moment operator $\bar{\mu}$. Reasoning by analogy, we can introduce an operator of "thermal" transitions $\hat{\tau}$ with matrix elements being essentially the probabilities of such processes. These processes lead to establishment of macroscopic thermodynamical equilibrium, which is an effect of interest here, and increase/decrease of temperature when kinetic thermal energy of molecules is being absorbed/released. Therefore, while not working out the structure and details, we introduce operator $\hat{\tau} = \hat{\tau}(R)$ as an operator of translational interaction in the ensemble of colliding molecules that necessarily results in maxwellian distribution of velocities. This operator acts on translational wavefunction $\Psi^T$ yielding $\Psi^{T'}$ so that the whole set $\{\Psi^{T'}\}$ for the ensemble of pairs corresponds to equilibrium distribution of velocities at temperature $T'$, and function $\Psi_{equ}^{T'}$ averaged over all pairs is related to mean parameters of thermal motion of molecules (in particular, to mean relative velocity $\bar{u}$). The dipole moment operator of two molecules reads as $\vec{\mu} = \vec{\mu}^A + \vec{\mu}^B$, where each $\vec{\mu}^A$ and $\vec{\mu}^B$ is a function of only internal coordinates of molecule A or B, respectively. For neutral molecules, even in case of strong interaction, the dipole moment operator $\vec{\mu}$ is an additive function and is independent of separation between molecules $R$. Therefore, the probability of optical transition between states $\Psi_1$ and $\Psi_2$ is defined only by the matrix elements of dipole moment which in zero approximation (5.1) gives $\vec{\mu}_{12} = \vec{\mu}_{12}^A \delta_{12}^B \delta_{12}^T + \vec{\mu}_{12}^B \delta_{12}^A \delta_{12}^T$. It is clear that this approximation explains optical transitions with a change of vibronic (internal) states of one of the molecules (A or B) only so that their translational state (and, hence, the temperature) is left invariable. Similarly, the "thermal" operator is $\tau_{12} = \tau_{12}^T \delta_{12}^A \delta_{12}^B$ and there are allowed "thermal" transitions without a change of internal molecular states. Consequently, in zero approximation the effects of optical and "thermal" energy exchange in gas are clearly separated: nonradiative transitions do not occur upon optical excitation, whereas "thermal" energy transfer due to collisions may take place (which is observable as a macroscopic effect).

To account for nonradiative transitions, we should refine the model by going to first approximation that includes dependencies of electronic parts of adiabatic wave functions upon translation coordinate $R$ $\left(\Psi_e^A(R), \Psi_e^B(R)\right)$ and mixing internal motions of nuclei $\left(\Psi_v^A(R)\Psi_v^B(R)\right)$ with translations $\left(\Psi^T\right)$. This can be done by solving the auxiliary vibrational problem for the system A+B with a very low force constant introduced for translation coordinate and additional nondiagonal matrix elements that describe mutual deformation of molecules upon close approach. The dependences $\Psi_e^A(R), \Psi_e^B(R)$ can be obtained by conventional quantum chemistry methods if one solve for polarization of electron shells in the field of partner molecule at varying distances between centers of mass of A and B.

So, the first approximation Hamiltonian can be presented in the form:

$$\hat{H} = \hat{H}_0 + \Delta\hat{H}_e^{A,B} + \Delta\hat{H}_{vT}^{A,B} \ , \tag{5.5}$$

where the zero approximation Hamiltonian $\hat{H}_0$ (5.1) is perturbed by mutual polarization of electron shells $\left(\Delta\hat{H}_e^{A,B}\right)$ and by interaction of internal vibrational coordinates with translation coordinate $R$ in the potential function of nuclei motion of molecular pair $\left(\Delta\hat{H}_{vT}^{A,B}\right)$.

Solution for the first approximation model (5.5) can be found by variational methods in the basis of zero approximation eigenfunctions assuming that their overlap matrix is unitary (the overlap between electron wavefunctions of molecules A and B in common Cartesian coordinates is negligible) and the diagonal elements of matrix $H$ are just total energies of the system in zero approximation. The nondiagonal elements will be then:

$$H_{kl} = \int \Psi_k \left( \Delta\hat{H}_e^{A,B} + \Delta\hat{H}_{vT}^{A,B} \right) \Psi_l dV =$$
$$= \int \Psi_{ek}^A \Psi_{vk}^A \Psi_{ek}^B \Psi_{vk}^B \Psi_k^T \Delta\hat{H}_e^{A,B} \Psi_{el}^A \Psi_{vl}^A \Psi_{el}^B \Psi_{vl}^B \Psi_l^T dV + \quad (5.6)$$
$$+ \int \Psi_{ek}^A \Psi_{vk}^A \Psi_{ek}^B \Psi_{vk}^B \Psi_k^T \Delta\hat{H}_{vT}^{A,B} \Psi_{el}^A \Psi_{vl}^A \Psi_{el}^B \Psi_{vl}^B \Psi_l^T dV.$$

In the simplest form of operator $\Delta\hat{H}_e^{A,B}$, when interaction between electron shells reduces to dipole-dipole interaction and depends only on mutual orientation of dipoles and distance between them, nondiagonal (with respect to vibronic wavefunctions of A and B) matrix elements $\left(\Delta\hat{H}_e^{A,B}\right)_{kl}$ vanish. Dipole-dipole interactions do not mix electron wavefunctions, no matter whether they belong to a single molecule or different molecules, i.e. purely electron wavefunctions do not change. Indeed, since each molecule is considered a point dipole and all the electrons are found in a uniform constant field, this gives only an additive correction to electron energy. Therefore, to this approximation, the only functions that mix are the vibrational wavefunctions of A and B and translational functions (through operator $\Delta\hat{H}_{vT}^{A,B}$). This is equivalent to vibrational relaxation, i.e. nonradiative (due to collisions) transformation of internal energy of vibrationally excited molecules into the energy of thermal motion accompanied by the change in temperature.

For the mixing of electron functions and, hence, for the electron relaxation to become significant, it is necessary to take into account mutual polarizations of electron shells and corresponding corrections to their wavefunctions. When the dependence of electron functions on mutual orientation of molecules is small, we can stick with the simple expansion:

$$\Psi_e^A(R) = \Psi_e^A(\infty) + \Delta\Psi_e^A \frac{1}{R} = \Psi_{e0}^A + \Delta\Psi_e^A \frac{1}{R} \quad (5.7)$$

(and for the molecule B in the same manner). Besides the method mentioned above, coefficients $\Delta\Psi_e^A$ in (5.7) can be found applying the theory of perturbations with a given form of polarization part of operator $\Delta\hat{H}_e^{A,B}$.

Taking (5.7) and discarding the second order terms, we obtain the nondiagonal elements (5.6)

$$H'_{kl} = \delta_{vkl}^A \delta_{vkl}^B \alpha_{kl} \int \Psi_k^T R^{-1} \Delta\hat{H}_e^{A,B} \Psi_l^T dR, \quad (5.8)$$

$$H''_{kl} = \delta_{ekl}^A \delta_{ekl}^B \int \Psi_{vk}^A \Psi_{vk}^B \Psi_k^T \Delta\hat{H}_{vT}^{A,B} \Psi_{vl}^A \Psi_{vl}^B \Psi_l^T dV, \quad (5.9)$$

where $\alpha_{kl} = \alpha_{kl}^A \delta_{ekl}^B + \alpha_{kl}^B \delta_{ekl}^A$, $\alpha_{kl}^A = \int \left[ \Psi_{e0k}^A \Delta\Psi_{el}^A + \Psi_{e0l}^A \Delta\Psi_{ek}^A \right] dV_e^A$ (the same procedure applies for B); and, by the orthonormal properties of basis functions, $\delta_{wkl}^C \equiv \int \Psi_{wk}^C \Psi_{wl}^C dV_w^C = \delta_{kl}$ ($\delta_{kl}$ is Kronecker delta-symbol; C=A,B; $w = e, v$). The matrix elements of the first kind $H'_{kl}$ are responsible for mixing electron and translational states (functions) without a change in vibrational functions; quantities $H''_{kl}$, in turn, describe mixing vibrational and translational functions leaving electronic parts unchanged. It is possible to write also the matrix elements that would mix electron, vibrational and translational states altogether

$$H'''_{kl} = \alpha_{kl} \int \Psi_{vk}^A \Psi_{vk}^B \Psi_k^T R^{-1} \Delta \hat{H}_{vT}^{A,B} \Psi_{vl}^A \Psi_{vl}^B \Psi_l^T dV , \qquad (5.10)$$

but these are smaller than previous two since they are proportional to $\alpha \Delta H_{vT}^{A,B}$.

Thus, operator $\Delta H_{vT}^{A,B}$ can be written as following:

$$\Delta H_{vT}^{A,B} = \sum_i u_{iR}^A q_i^A R + \sum_j u_{jR}^B q_j^B R , \qquad (5.11)$$

where $q_i^A$, $q_j^B$ represent natural vibrational coordinates of molecules A and B; $u_{iR}^A$, $u_{jR}^B$ are the force constants attributed to interaction between internal motions with translation coordinate. We omit here the term $u_R R^2$ along with the matrix elements of the Hamiltonian that are proportional to $\delta_{ekl}^A \delta_{ekl}^B \delta_{vkl}^A \delta_{vkl}^B \int \Psi_k^T \Delta \hat{H}_e^{A,B} \Psi_l^T dR$ reasoning that these factors have already been accounted for in model Morse potential (and in Hamiltonian $H_0$) which describes translations in zero approximation. The integrals (5.9) can then be easily computed since these quantities reduce to matrix elements of harmonic oscillator $\int \Psi_{vk} q \Psi_{vl} dq$ and Morse oscillator $\int \Psi_k^T R \Psi_l^T dR$ or $\int \Psi_k^T R^{-n} \Psi_l^T dR$ for (5.8).

Small nondiagonal elements of variational matrix $H$ (see (5.8), (5.9) and (5.10)) may result in significant mixing between basis functions only when corresponding zero approximation energy levels are in resonance. Under such conditions, population of vibronic states excited due to optical dipole transitions will be accompanied by population of translational state with big increment in kinetic energy of relative motion of molecules A and B. This, in turn, will lead to excess number of fast molecules in the medium (compared to equilibrium distribution at temperature $T_0$). Collisions will dissipate these fluctuations into new equilibrium state of the gas at higher temperature $T_1 > T_0$. So, part of absorbed radiation energy will be converted to thermal energy of molecular motion.

Consider this in more detail. Let's assume that the system is in thermal equilibrium at temperature $T' = T_0$ and all $N$ pairs of molecules are in averaged $\Psi'$ state which is essentially the combination of molecular ground states $\Psi'^A, \Psi'^B$ and translational state $\Psi'^T$ of thermal motion with mean velocity $\bar{u}' = 4\sqrt{kT'/\pi M}$. This means that all pairs are treated indistinguishable with respect to their translational dynamics averaged over statistical ensemble. As noted above, this allow to introduce the constraint $0 < R < \lambda$ for translations and define function $\Psi'^T$ at temperature $T'$.

To not complicate the discussion, we will assume that initial states do not mix and write them in zero approximation:

$$\Psi' = \Psi'^A \Psi'^B \Psi'^T. \qquad (5.12)$$

Suppose further that only molecule A is initially excited and take the mix of two resonant energy levels:

$$\Psi'' = c_1 \left( \Psi''^A \Psi'^B \Psi'^T \right) + c_2 \left( \Psi'^A \Psi'^B \Psi''^T \right). \qquad (5.13)$$

Then optical excitation considered in zero approximation will leave the same maxwellian distribution at temperature $T'$ unchanged.

Going to first approximation we note that molecules will undergo transition into states with mixed wavefunction (5.13). Each $i$-th pair with relative velocity $u_i$ will be excited into state $\Psi''_i$

$$\Psi''_i = c_1^i \left( \Psi''^A \Psi'^B \Psi'^T_i \right) + c_2^i \left( \Psi'^A \Psi'^B \Psi''^T_i \right) \qquad (5.14)$$

(with initial state $\Psi'_i = \Psi'^A \Psi'^B \Psi'^T_i$). The probability of optical excitation for each pair is proportional to the dipole moment squared

$$\bar{\mu}_i^{',\,''} = \langle \Psi' | \vec{\mu} | \Psi''_i \rangle = c_1^i \bar{\mu}_A^{',\,''}, \qquad (5.15)$$

where $\bar{\mu}_A^{',\,''} = \int \Psi''^A \vec{\mu} \Psi'^A dV$, i.e. shows no explicit dependence upon translational states $\Psi'^T_i$, but do depend on mixing coefficients $c_1^i$ that vary for different pairs (see below). The total number of optically excited pairs $\Delta N$ (population of mixed state) will be in proportion to the total probability:

$$\left( \bar{\mu}^{',\,''} \right)^2 = \sum_i^N \left( \bar{\mu}_i^{',\,''} \right)^2 = \left( \bar{\mu}_A^{',\,''} \right)^2 \sum_i^N \left( c_1^i \right)^2. \qquad (5.16)$$

In general, the velocity distribution for this statistical subsystem $\Delta N$ will deviate from maxwellian distrubution since the number of pairs moving with a given velocity $u_i$ will be proportional to the product $N_u \left( c_1^i \right)^2$ ($N_u$ is the initial Maxwell distribution), i.e. it will depend on $\left( c_1^i \right)^2$.

The mixing coefficients $c_1^i$ (as well as $c_2^i$) originate from nondiagonal elements of variational matrix (5.8), (5.9) and (5.10) and, hence, from integrals of the form $\int \Psi'^T R^n \Psi''^T dR$ which, using the approximate representation of Morse functions, are

$$\int \Psi'^T R^n \Psi''^T dR \approx N'N'' \int f'(R-R_0)f''(R-R_0)R^n Cos\omega'(R-R_0)Cos\omega''(R-R_0)dR =$$

$$= \frac{N'N''}{2}\int f'(R-R_0)f''(R-R_0)R^n[Cos(\omega'-\omega'')(R-R_0)+Cos(\omega'+\omega'')(R-R_0)]dR = \quad (5.17)$$

$$= \varphi_1(\omega'-\omega'')+\varphi_2(\omega'+\omega'') \approx \varphi_1(\omega'-\omega''),$$

because $\varphi_2(\omega'+\omega'') \ll \varphi_1(\omega'-\omega'')$ due to the presence of fast oscillatory factor $Cos(\omega'+\omega'')R$ in the integrand (5.17).

Note that as long as resonance conditions hold (see above), the change in translational kinetic energy $\Delta E_T$ is the same for all pairs and it is also the energy of vibronic excitation $\Delta E_T = \Delta E_{ev}$ itself. Therefore, $\omega'-\omega'' = const$ and, in the first approximation, all the coefficients $c_1^i$ can be assumed equal $c_1^i = c_1$. Correspondingly,

$$(\vec{\mu}^{',\,''})^2 = (\vec{\mu}_A^{',\,''})^2 c_1^2 N \qquad (5.18)$$

and velocity distribution of the excited subsystem $\Delta N$ will be again equilibrium (maxwellian), but with different mean velocity.

Indeed, for the mean-square velocity $u'' = \sqrt{\langle u''^2 \rangle}$ of the subsystem $\Delta N$ we have:

$$\langle u''^2 \rangle = \int \Psi'' u^2 \Psi'' dV = c_1^2 \langle u^2 \rangle' + c_2^2 \langle u^2 \rangle'' = \langle u^2 \rangle' + c_2^2 \Delta u^2, \qquad (5.19)$$

where $\langle u^2 \rangle' = \int \Psi'^T u^2 \Psi'^T dR = \langle u'^2 \rangle = u_0^2$ ( $u_0$ is the mean-square velocity in the initial state), $\langle u^2 \rangle'' = \int \Psi''^T u^2 \Psi''^T dR$, $\Delta u^2 = \langle u^2 \rangle'' - \langle u^2 \rangle'$ and it is kept in mind that $c_1^2 + c_2^2 = 1$ by normalization. The subsystem of "internally excited" pairs of molecules $\Delta N$ $(\Psi'^A \to \Psi''^A)$ will have its own temperature $T'' \ne T'$ and, since $T \sim \langle u^2 \rangle$, we get:

$$T'' = T'\frac{u''^2}{u_0^2} = T'\left[1+c_2^2\frac{\Delta u^2}{u_0^2}\right] > T'. \qquad (5.20)$$

Thus, as a result of optical excitation, the gas consisting of $N$ pairs of molecules undergoes transition to thermodynamically nonequilibrium state which is a composition of two statistically "equilibrium" subsystems: $(N-\Delta N)$ pairs of unexcited molecules at temperature $T'$ with mean-square velocity $u_0$; and $\Delta N$ pairs of excited molecules with $T''$ and $u''$, respectively.

Collisions modeled by operator $\hat{\tau}$ bring the system to thermodynamical equilibrium at new (finite) temperature $T_1$. Presuming that in the first approximation operator $\hat{\tau}$ does not affect internal molecular states and the total energy of ensemble is conserved, we obtain with (5.19) that

$$N\frac{Mu_1^2}{2} = \Delta N\frac{Mu''^2}{2}+(N-\Delta N)\frac{Mu_0^2}{2}, \quad u_1^2 = u_0^2 + \frac{\Delta N}{N}c_2^2\Delta u^2. \text{ Since } \Delta N/N \sim (\vec{\mu}^{',\,''})^2/N = (\vec{\mu}_A^{',\,''})^2 c_1^2$$

then $\dfrac{u_1^2}{u_0^2} = 1 + \sigma(\vec{\mu}_A{'}{,}{"})^2 c_1^2 c_2^2 \dfrac{\Delta u^2}{u_0^2}$ and $\dfrac{T_1}{T_0} = 1 + \sigma(\vec{\mu}_A{'}{,}{"})^2 c_1^2 c_2^2 \dfrac{\Delta u^2}{u_0^2}$, where $\sigma$ is a coefficient in the expression for optical transition probability $w = \sigma(\vec{\mu}{'}{,}{"})^2$ which depend, in particular, on radiation flux density. Taking into account that for the model of pair-wise interactions $E_T = \dfrac{1}{2}kT$ and also $\Delta E_T = \Delta E_{ev}$ from resonance conditions, it follows that $\dfrac{\Delta u^2}{u_0^2} = \dfrac{\delta T}{T_0} = 2\dfrac{\Delta E_T}{kT_0} = 2\dfrac{\Delta E_{ev}}{kT_0}$ and

$\dfrac{T_1}{T_0} = 1 + 2\sigma(\vec{\mu}_A{'}{,}{"})^2 c_2^2 \dfrac{\Delta E_{ev}}{kT_0} - 2\sigma(\vec{\mu}_A{'}{,}{"})^2 c_2^4 \dfrac{\Delta E_{ev}}{kT_0}$ where we neglect the second order in $c_2^2$ terms (i.e. those that are proportional to $c_2^4$).

The first approximation model shows thereby that after absorption of radiation the temperature of gas will be $T_1 = T_0\left[1 + 2\sigma(\vec{\mu}_A{'}{,}{"})^2 c_2^2 \dfrac{\Delta E_{ev}}{kT_0}\right]$. The increment in temperature due to optical excitation is $\Delta T = T_1 - T_0 = 2\sigma(\vec{\mu}_A{'}{,}{"})^2 c_2^2 \dfrac{\Delta E_{ev}}{k}$ and the ratio of gained thermal energy ($\Delta E_{therm} = N\dfrac{1}{2}k\Delta T$) to energy deposited through radiation ($\Delta E_{abs} = w\Delta E_{ev}$) is then

$$\Delta E_{therm}/\Delta E_{abs} = \dfrac{1}{2}k\Delta T \Big/ w\Delta E_{ev} = c_2^2 \ . \tag{5.21}$$

Expression (5.21) holds for $c_2^2 \ll c_1^2$ and $c_2^2 \sim c_1^2$ as well. In particular, when mixing of wavefunctions is complete (degenerate energy levels), $c_1^2 = c_2^2 = 1/2$, $\Delta E_{therm}/\Delta E_{abs} = 1/2$ and, hence, a half of radiation energy will be converted into heat.

Here we see how the model of molecular interactions describes nonradiative transformation of absorbed energy into thermal energy of chaotic motion. As a consequence, the fluorescence quantum yield will be less than unity, which is observed in the experiments. The proposed approach enables to make quantitative estimates for nonradiative effects, since all matrix elements and coefficients in model operators can be easily calculated. Note that other cases of mixing between molecular states can be examined in a similar way. Consider, for example, states with wavefunctions $\Psi''^A\Psi'^B\Psi'^T$ and $\Psi'''^A\Psi'^B\Psi''^T$ $\left(\Psi''^A = \Psi_e''^A\Psi_v''^A,\ \Psi'''^A = \Psi_e''^A\Psi_v'''^A\right)$ that correspond to one and the same excited electron state of molecule A $\left(\Psi_e''^A\right)$ and various vibrational sublevels $\left(\Psi_v''^A\ \text{and}\ \Psi_v'''^A\right)$; or various excited electron states of molecule A. Apparently, the first case gives an approach to nonradiative vibrational relaxation of excited vibronic state, while the second deals with electron relaxation. Both types of relaxation can be observed experimentally and described qualitatively by Vavilov's law and Kasha's rule, respectively.

Despite the fact that we have used restricted "two-level" approximation for simplicity, the method is applicable to "multilevel" mixing due to complex molecular interactions too. This, however, implies diagonalization of full variational matrix, which in no way represents a serious computational problem even for a big number of mixing vibronic states and high dimensional matrices [93]. It is also noteworthy for future studies and possible applications that the model of interactions between molecules in gases considered herein can be naturally incorporated into methods and software developed for simulations of dynamical time-resolved spectra.

**Conclusions**

We have considered recent advances and applications in the adiabatic semiempirical theory of vibronic spectra, along with major features of the parametric method and associated computational techniques that now extend to time-resolved studies of polyatomic molecules, isomer transformations and nonradiative processes. Indeed, extensive computer simulations and approaches to inverse problems will be expected to fill the gap between purely theoretical molecular models and increasingly sophisticated spectral experiments, which, in any case, can give only indirect information on properties and dynamics of excited states. It is also clear that current models should eventually give way to more accurate approximations (for example, nonadiabatic theory, full treatment of time-dependent problem, etc.), but there will still be a challenge for maintaining intuitive description and clarity.

**Acknowledgements**

One of the authors (S.A.) thanks professor David Farrelly at Utah State University for reading the first version of the manuscript, valuable suggestions and support all the way through. Partial financial support of this work by the Russian Foundation for Basic Research (project No. 01-03-32058) and the grant for scientific schools (No. НШ-1186.2003.3) is also acknowledged.

**References**

1. M. Tian, F. Grelet, I. Lorgere, J.-P. Galaup, J.-L. le Gouet. *J. Opt. Soc. Am. B.*, **16** (1) (1999) 74.
2. Collision-Based Computing, Ed.: A. Adamatzky (Springer-Verlag, 2002), XXVII.
3. T.D. Schneider. *Nanotechnology*, **5** (1994) 1.

4. T.D. Schneider. *J. Theor. Biol.*, **148** (1991) 83.
5. V. Balzani. *Photochem. Photobiol. Sci.*, **2** (5) (2003) 459.
6. A.P. Davis. *Nature*, **401** (1999) 120.
7. C. Joachim, J. K. Gimzewski, A. Aviram. *Nature,* **408** (2000) 541.
8. D. Philp, J.F. Stoddart. *Angewandte Chemie - International Edition*, **35** (11) (1996) 1155.
9. M. Schliwa, G. Woehlke. *Nature,* **422** (2003) 759.
10. L.A. Gribov. Intensity Theory for Infrared Spectra of Polyatomic Molecules. Consultant Bureau, New York, 1964.
11. M.V. Volkenstein, L.A. Gribov, M.A. Eliashevich, B.I. Stepanov. Molecular Vibrations. Main Editorial Board for Physico-Mathematical Literature, Moscow, 1972 (in Russian).
12. M.E. Elyashberg, L.A. Gribov, V.V. Serov. Molecular Spectral Analysis and Computer. Nauka, Moscow, 1980 (in Russian).
13. L.A. Gribov, V.A. Dementiev. Computational Methods and Algorithms in the Theory of Molecular Vibrational Spectra. Nauka, Moscow, 1981 (in Russian).
14. L.A. Gribov, V.I. Baranov, B.K. Novosadov. Methods of Computing Electronic-Vibrational Spectra of Polyatomic Molecules. Nauka, Moscow, 1984 (in Russian).
15. L.A. Gribov, W.J. Orville-Thomas. Theory and Methods of Calculation of Molecular Spectra. Wiley, New York, 1988.
16. L.A. Gribov, V.I. Baranov, D.Yu. Zelent'sov. Electronic-Vibrational Spectra of Polyatomic Molecules: Theory and Methods of Calculation. Nauka, Moscow, 1997 (in Russian).
17. L.A. Gribov, V.I. Baranov, M.E. Elyashberg. Standardless Molecular Spectral Analysis. Theoretical Foundations. Editorial URSS, Moscow, 2002 (in Russian).
18. V.I. Baranov, L.A. Gribov, B.K. Novosadov. *J. Mol. Struct.*, **70** (1981) 1.
19. V.I. Baranov, D.Yu. Zelent'sov. *J. Mol. Struct.*, **328** (1994) 179.
20. V.I. Baranov, L.A. Gribov, D.Yu. Zelent'sov. *J. Mol. Struct.*, **376** (1996) 475.
21. V.I. Baranov, L.A. Gribov. *J. Mol. Struct.*, **70** (1981) 31.
22. V.I. Baranov, A.N. Solov'ev. *Zhurnal Phizicheskoy Khimii*. **59** (1985) 1720 (in Russian).
23. V.I. Baranov, G.N. Ten, L.A. Gribov. *J. Mol. Struct.*, **137** (1986) 91.
24. V.I. Baranov, A.N. Solov'ev. *Optika i Spektroskopiya (Opt. Spectrosc.)*, **62** (1987) 346 (in Russian).
25. V.I. Baranov. *J. Appl. Spectr.*, **51** (1989) 842.
26. M. Ito, T. Ebata, N. Mikami. Laser Spectroscopy of Large Polyatomic Molecules in Supersonic Jets. *Ann. Rev. Phys. Chem.*, **39** (1988) 123.
27. J.A. Syage, P.M. Felker, D.H. Semmes, F.Al. Adel, A.H. Zewail. *J. Chem. Phys.*, **82** (1985) 2896.

28. O. Kaimoto, S. Hayami, H. Shizuka. *Chem. Phys. Lett.*, **177** (1991) 219.
29. C.M. Cheatham, M. Fluang, N. Meinander, M.B. Kelly, K. Haler, W.Y. Chiang, J. Laane. *J. Mol. Struct.*, **377** (1996) 81.
30. S. Mukamel. *J. Chem. Phys.*, **82** (1985) 2867.
31. G.N. Patwari, S. Doraiswami, S. Wategaonkar. *Chem. Phys. Lett.*, **316** (2000) 433.
32. G.N. Patwari, S. Doraiswamy, S. Wategaonkar. *Chem. Phys. Lett.*, **305** (1999) 381.
33. J.A. Syage, P.M. Felker, A.H. Zewail. *J. Chem. Phys.*, **81** (1984) 4685.
34. J. Prochorow, I. Deperasinska, O. Morawski. *Chem. Phys. Lett.*, **316** (2000) 24.
35. G.S. Harms, T. Irngartinger, D. Reiss, A. Renn, U.P. Wild. *Chem. Phys. Lett.*, **313** (1999) 533.
36. A.A. Heikal, J.S. Baskin, L. Banares, A.H. Zewail. *J. Phys. Chem. A.*, **101** (1997) 572.
37. K. Tsuji, C. Terauchi, K. Shibuya, S. Tsuchiya. *Chem. Phys. Lett.*, **306** (1999) 41.
38. G.N. Patwari, S. Wategaonkar. *Chem. Phys. Lett.*, **323** (2000) 460.
39. Femtochemistry and Femtobiology. Eds: A. Douhal, J. Santamaria, World Scientific Pub Co; 1st edition, 2002.
40. A.H. Zewail, Femtochemistry, *Adv. in Chem. Phys.*, **101** (1997) 892.
41. Femtochemistry: Ultrafast Chemical and Physical Processes in Molecular Systems, Ed.: M. Chergui. World Scientific, Singapore, 1996.
42. A.H. Zewail, Femtochemistry − Ultrafast Dynamics of the Chemical Bond, Vol. I and II, World Scientific (20th Century Chemistry Series), New Jersey, Singapore, 1994.
43. A.H. Zewail. Femtochemistry: Atomic-Scale Dynamics of the Chemical Bond Using Ultrafast Lasers - (Nobel Lecture), *Angewandte Chemie - International Edition*, **39** (15) (2000) 2587.
44. E.D. Potter, J.L. Herek, S. Pedersen, Q. Liu, and A.H. Zewail, *Nature*, **355** (1992) 66.
45. A.H. Zewail. *Angewandte Chemie - International Edition*, **40** (23) (2001) 4371.
46. A.H. Zewail. *Science*, **242** (4886) (1988) 1645.
47. V.I. Baranov, L.A. Gribov. *J. Appl. Spectr.*, **48** (1988) 629.
48. V.I. Baranov, L.A. Gribov, V.O. Djenjer, D.Yu. Zelent'sov. *J. Mol. Struct.*, **407** (1997) 177.
49. V.I. Baranov, L.A. Gribov, V.O. Djenjer, D.Yu. Zelent'sov. *J. Mol. Struct.*, **407** (1997) 199.
50. V.I. Baranov, L.A. Gribov, V.O. Djenjer, D.Yu. Zelent'sov. *J. Mol. Struct.*, **407** (1997) 209.
51. V.I. Baranov, A.N. Solov'ev. *Zhurnal Strukturnoi Khimii (J. Struct. Chem.)*, **41** (2000) 368 (in Russian).
52. V.I. Baranov, A.N. Solov'ev. *Opt. Spectrosc.*, **90** (2001) 183.
53. V.I. Baranov, A.N. Solov'ev. *Opt. Spectrosc.*, **93** (2002) 690.
54. V.I. Baranov. *J. Appl. Spectr.*, **51** (1989) 1072.

55. L.A. Gribov, V.I. Baranov. *Zhurnal Prikladnoi Spektroscopii (J. Appl. Spectr.)*, **44** (1986) 341 (in Russian).
56. L.A. Gribov, V.I. Baranov. *J. Mol. Struct.*, **104** (1983) 267.
57. V.I. Baranov, L.A. Gribov. *Optika i Spectroskopiya (Opt. Spectrosc.)*, **67** (1989) 32 (in Russian).
58. L.A. Gribov, V.I. Baranov. *J. Mol. Struct.*, **224** (1990) 45.
59. P. Pulay. *Mol. Phys.*, **18** (1970) 473; P. Pulay, F. Torok. *Mol. Phys.*, **25** (1973) 1153; G. Fogarasi, P. Pulay. *Ann. Rev. Phys. Chem.*, **35** (1984) 191; G.J. Sexton, N.C. Handy. *Mol. Phys.*, **51** (1984) 1321; G. Fogarasi, P. Pulay. *J. Mol. Struct.*, **141** (1986) 145.
60. V.S. Fikhtengoltz, R.V. Zolotareva, Yu.A. Lvov. Atlas of Ultraviolet Absorption Spectra of Substances Used in Synthetic Caoutchouc Manufacture. Khimiya, Leningrad, 1969 (in Russian).
61. D.G. Leopold, R.D. Pendley, J.L. Roebber, R.J. Hemley, V.J. Vaida. *Chem. Phys.*, **81** (1984) 4218.
62. S.A. Cosgrove, M.A. Guite, T.B. Burnell, R.J. Christensen. *J. Chem. Phys.*, **94** (1990) 8118.
63. D.G. Leopold, V. Vaida, M.F. Granville. *J. Chem. Phys.*, **81** (1984) 4210.
64. W.G. Bouwman, A.C. Jones, D. Phillips, P. Thibodeau, C. Friel, R.L. Christensen. *J. Phys. Chem.*, **94** (1990) 7429.
65. V.I. Baranov. *Zhurnal Prikladnoi Spektroscopii (J. Appl. Spectr.)*, **67** (2000) 148 (in Russian).
66. R.J. Hemley, A.C. Lasaga, V. Vaida, M. Karplus. *J. Phys. Chem.*, **92** (1988) 945.
67. W.R. Lambert, P.M. Felker, J.A. Syage, A.H. Zewail. *J. Chem. Phys.*, **81** (1984) 2195.
68. T.N. Bolotnikova, V.A. Zgukov, L.V. Utkina, V.I. Shaposhnikov. *Optika i Spectroskopiya (Opt. Spectrosc.)*, **53** (1982) 823 (in Russian).
69. A. Amirav, U. Even, J.J. Jortner. *Chem. Phys.*, **75** (1981) 3770.
70. V.I. Baranov, L.A. Gribov. *Zhurnal Prikladnoi Spektroscopii (J. Appl. Spectr.)*, **67** (2000) 289 (in Russian).
71. L.A. Gribov, I.V. Maslov. *J. Mol. Struct.*, **521** (2000) 107.
72. L.A. Gribov, V.I. Baranov. *Optika i Spectroskopiya (Opt. Spectrosc.)*, **85** (1998) 46 (in Russian).
73. V.I. Baranov, L.A. Gribov, D.Yu. Zelent'sov. *J. Mol. Struct.*, **328** (1994) 189.
74. V.I. Baranov, D.Yu. Zelent'sov. *J. Mol. Struct.*, **328** (1994) 199.
75. V.I. Baranov, F.A. Savin, L.A. Gribov. Programs for Calculation of Electronic-Vibrational Spectra of Polyatomic Molecules. Nauka, Moscow, 1983 (in Russian).
76. S.A. Astakhov, V.I. Baranov. *Opt. Spectrosc.*, **90** (2001) 199.
77. V.I. Arnold. Ordinary differential equations. MIT Press, Cambridge, Massachusetts, 1973.
78. J.A. Syage, P.M. Felker, A.H. Zewail. *J. Chem. Phys.*, **81** (1984) 4706.
79. E.N. Borisov, L.P. Kantserova, V.B. Smirnov, V.Yu. Cherepanov. *Opt. Spectrosc.*, **91** (2001) 900.
80. G.N. Dovzhenko, D.L. Dovzhenko, V.B. Smirnov. *Opt. Spectrosc.*, **86** (1999) 20.

81. L.A. Gribov, V.I. Baranov, S.A. Astakhov. Doklady Chemistry, **374**(4) (2000) 203; S.A. Astakhov, V.I. Baranov, L.A. Gribov. *J. Analyt. Chem.*, **56** (2001) 625.
82. S.A. Astakhov, V.I. Baranov, L.A. Gribov. *J. Mol. Struct.*, **655** (2003) 97.
83. S.E. Bialkowski. Photothermal Spectroscopy Methods for Chemical Analysis. Volume 134 in Chemical Analysis, Wiley, New York, 1996.
84. S. Yamamoto, K. Okuyama, N. Mikami, M. Ito. *Chem. Phys. Lett.*, **125** (1986) 1.
85. J. Saltiel, Y. Zhang, D.F. Sears Jr. *J. Phys. Chem. A.*, **101** (1997) 7053.
86. V.I. Baranov. *Opt. Spectrosc.*, **88** (2000) 182.
87. S.A. Astakhov, V.I. Baranov. *Opt. Spectrosc.*, **92** (2002) 20.
88. L.A. Gribov. From Theory of Spectra to Theory of Chemical Transformations. Editorial URSS, Moscow, 2001 (in Russian).
89. B.I. Stepanov. *Uspekhi Fizicheskikh Nauk (Adv. Phys. Sci.)*, **58** (1956) 3 (in Russian).
90. M. Kasha. *Discuss. Faraday Soc.*, **9** (1950) 14.
91. L.A. Gribov, I.V. Maslov. *Zhurnal Phizicheskoy Khimii.* **74** (3) (2000) 441 (in Russian).
92. B.N. Zakhar'ev. Lessons in Quantum Intuition. JINR, Dubna, 1996 (in Russian).
93. L.A. Gribov, V.I. Baranov, Yu.V. Nefedov. *J.Mol. Struct.*, **148** (1986) 1.